\documentclass[12pt]{article}
\usepackage[utf8]{inputenc}
\usepackage[british]{babel}
\usepackage{cmap}
\usepackage{lmodern}

\usepackage{amssymb, amsmath, amsthm}
\usepackage[a4paper,top=25mm,bottom=25mm,left=25mm,right=25mm]{geometry}
\usepackage{ragged2e}

\usepackage{authblk} % for headings
\usepackage{pifont}
\usepackage{graphicx}
\usepackage[dvipsnames,svgnames,table]{xcolor}
\usepackage[figuresright]{rotating}
\usepackage{xtab} % tackle the long tables
\usepackage{longtable} % tackle the long tables
\usepackage{multirow}
\usepackage{footnote}
\usepackage[stable]{footmisc}
\usepackage{chngpage} % allows for temporary adjustment of side margins
\usepackage{pdflscape} % landscape environment
\usepackage[nottoc,notlot,notlof]{tocbibind} % includes everything in ToC
\usepackage{appendix}
\usepackage{graphicx}

\usepackage{pgfplots}
\pgfplotsset{compat=1.18}
\pgfplotsset{every tick label/.append style={font=\footnotesize}}
\usepackage{setspace}
\makesavenoteenv{tabular}
\usepackage{tabularx}
\usepackage{booktabs}
\usepackage{threeparttable} % [flushleft]
\usepackage[referable]{threeparttablex} % footnotes in tabu
\newcolumntype{R}{>{\raggedleft\arraybackslash}X}
\newcolumntype{L}{>{\raggedright\arraybackslash}X}
\newcolumntype{C}{>{\centering\arraybackslash}X}
\newcolumntype{A}{>{\columncolor{gray!25}}C}
\newcolumntype{a}{>{\columncolor{gray!25}}c}

\newlength{\tablen}

\usepackage{dcolumn} % alignment to decimal points
\newcolumntype{.}{D{.}{.}{-1}}

\usepackage{tikz}
\usetikzlibrary{arrows, calc, matrix, patterns, positioning, shapes, trees}
\usepackage[semicolon]{natbib}
\usepackage[hyphens]{url}
\usepackage{hyperref} % [hidelinks]
\hypersetup{
  colorlinks   = true,    % Colours links instead of ugly boxes
  urlcolor     = blue,    % Colour for external hyperlinks
  linkcolor    = blue,    % Colour of internal links
  citecolor    = ForestGreen      % Colour of citations
}
\usepackage{microtype}
\usepackage[singlelinecheck=false,justification=centering]{caption} % [justification=centering]

\usepackage[labelformat=simple]{subcaption}
\makeatletter
\renewcommand{\thesubfigure}{(\alph{subfigure})}
\renewcommand{\p@subfigure}{\thefigure}
\makeatother

\DeclareCaptionLabelFormat{parenthesis}{(#2)}
\makeatletter
\renewcommand\p@subtable{\arabic{table}.}
\makeatother

\usepackage{enumitem}
\setlist[itemize]{leftmargin=2.5\parindent}
\setlist[enumerate]{leftmargin=2.5\parindent}

\def\addlegendimage{\csname pgfplots@addlegendimage\endcsname}

\theoremstyle{plain}
\theoremstyle{definition}
\theoremstyle{remark}

\def\keywords{\vspace{.5em} % Add keywords
{\noindent \textit{Keywords}: }}

\def\JEL{\vspace{.5em} % Add keywords
{\noindent \textbf{\emph{JEL} classification number}: }}

\def\AMS{\vspace{.5em} % Add keywords
{\noindent \textbf{\emph{MSC} class}: }}

\title{Encouraging offensive play: A hidden benefit \\ of incomplete round-robin tournaments}

\author{\href{https://sites.google.com/view/laszlocsato}{L\'aszl\'o Csat\'o}\thanks{~Corresponding author. Email: \emph{laszlo.csato@sztaki.hun-ren.hu} \newline
Institute for Computer Science and Control (SZTAKI), Hungarian Research Network (HUN-REN), Laboratory on Engineering and Management Intelligence, Research Group of Operations Research and Decision Systems, Budapest, Hungary \newline
Corvinus University of Budapest (BCE), Institute of Operations and Decision Sciences, Department of Operations Research and Actuarial Sciences, Budapest, Hungary}
$\qquad \qquad$
David Winkelmann\thanks{~Email: \emph{david.winkelmann@univie.ac.at} \newline
Department of Business Decisions and Analytics, University of Vienna, Vienna, Austria}}

\date{\today}

\def\Dedication{
{\noindent
``\emph{It is a commonplace among economists to hold up sports as an example of contest/tourna\-ment theory in action, but in practice a lot remains to be done both to understand the relationship between tournament structures and incentives in theory, and to test theories against the data.}''
\citep[p.~1181]{Szymanski2003}}
}
\vspace{0.5cm}

\begin{document}

\newgeometry{top=25mm,bottom=25mm,left=25mm,right=25mm}

\maketitle
\thispagestyle{empty}
\Dedication
\begin{abstract}
\noindent
The Union of European Football Associations (UEFA) fundamentally reformed its club competitions from the 2024/25 season onward. The traditional group stage has been replaced by an incomplete round-robin league phase, in which all teams are ranked in a single table based on matches played against a subset of the other teams. Early evidence from the first two seasons of the UEFA Champions League in the new format indicates a marked decline in the proportion of drawn matches compared to the old format.
Inspired by this observation, our paper examines how tournament design shapes the incentives to play offensively. We develop a simulation framework to compare the marginal gain from scoring a goal and loss from conceding a goal in terms of the probability of reaching critical ranking thresholds, using the initial scoreline of 0-0 as the reference point. We find robustly stronger incentives for offensive play in the new Champions League design. Roughly half of the increase is attributable to the modified thresholds, but the other half stems directly from replacing the group stage with an incomplete round-robin league. Our study is the first demonstration in the literature that the tournament format can materially affect team behaviour on the field.

\keywords{OR in sports; incentives; incomplete round-robin; simulation; tournament design; UEFA Champions League}

\AMS{62P20, 90-10, 90B90, 91B14}
% Applications of statistics to economics
% Mathematical modeling or simulation for problems pertaining to operations research and mathematical programming
% Case-oriented studies in operations research
% Social Choice

\JEL{C44, C53, D71, Z20}
% Operations Research, Statistical Decision Theory
% Simulation Methods
% Social Choice, Clubs, Committees, Associations
% Sports Economics, General
\end{abstract}

\clearpage
\restoregeometry

\section{Introduction} \label{Sec1}
%\label{sec:introductio}
%\input{files_DW/01_Introduction_DW}

%\citet{UEFA2024c}
%The new league format will ensure that any result has the potential to dramatically change a team's position, right up to and including the very last matchday.
Encouraging offensive play is one of the greatest challenges for tournament organisers in several sports. Perhaps the most substantial change in the recent history of football, the introduction of three points for a win rather than two, was explicitly intended to reduce the likelihood of draws and incentivise teams to attack \citep{LentenKendall2022}.
%The President of the Union of European Football Associations (UEFA), \emph{Aleksander {\v C}eferin}, has also explained the abolition of the away-goals rule in 2021 as follows \citep{UEFA2021f}: ``\emph{The impact of the rule now runs counter to its original purpose as, in fact, it now dissuades home teams -- especially in first legs -- from attacking, because they fear conceding a goal that would give their opponents a crucial advantage.}''

Naturally, decision-makers have various tools to alter the incentives of the contestants \citep{Medcalfe2026, Medina2026}. However, while behavioural responses to ranking (see Section~\ref{Sec2}) and prize-distribution rules \citep{EhrenbergBognanno1990, KondratevIanovskiNesterov2024} have been thoroughly analysed, the impact of the tournament format has not been studied yet from this perspective. A potential reason is the lack of empirical data, as significant changes in tournament formats are relatively rare.

The two basic competition formats are the knockout and the round-robin \citep{DevriesereCsatoGoossens2025}. The so-called \emph{incomplete round-robin} contest is increasingly used as well \citep{DevriesereGoossens2025}. Here, the teams compete in a single league similar to the standard round-robin, but they play only against a subset of the other teams. In contrast to the Swiss system \citep{Csato2013a, Csato2017a, DongRibeiroXuZamoraMaJing2023, SauerCsehLenzner2024}, the match schedule is fixed before the start of the tournament.

To the best of our knowledge, the first incomplete round-robin tournament in football was the qualifying phase of the 2019/20 CONCACAF Nations League. 34 teams were divided into four pots, and each team played two home and two away matches against four teams drawn from different pots. The Romanian football cup (Cupa Rom{\^ a}niei) has a stage with four groups since the 2022/23 season: a group consists of two teams from each of the three seeding pots, and each team plays three matches against teams from different pots in their group. The league phase of the North American Leagues Cup is contested by 18 Major League Soccer (MLS) and 18 Mexican clubs. Since 2025, the teams have been divided into six sets of six teams; these sets contain one team per league from each of the three tiers in two geographical regions, with these teams meeting the three foreign teams. For both leagues, a single table is constructed to rank the 18 teams.

The incomplete round-robin format has gained prominence primarily due to a recent decision of the Union of European Football Associations (UEFA). The reform, effective from the 2024/25 season, has replaced the traditional group stage of UEFA club competitions with an incomplete round-robin league phase. As Section~\ref{Sec2} details, the new format raises several interesting research questions related to the competitiveness of matches, the draw, the ranking, and the schedule that have been studied by recent operational research papers.%competitiveness of matches \citep{CsatoGyimesi2026b, DevriesereGoossensSpieksma2026, Gyimesi2024}, the draw \citep{CsatoGyimesiGoossensDevriesereLambersSpieksma2026, GuyonBenSalemBuchholtzerTanre2026, WesselSpieksma2026}, the ranking \citep{CsatoDevriesereGoossensGyimesiLambersSpieksma2026}, and the schedule \citep{DevriesereGoossensSpieksma2027, deWerraUrrutiaAssuncao2026, GuyonBenSalemBuchholtzerTanre2026}.

\begin{figure}[t!]
\centering

\begin{tikzpicture}
\begin{axis}[
width = \textwidth, 
height = 0.5\textwidth,
%title = Unweighted rounds,
%title style = {font=\small},
xmajorgrids = true,
ymajorgrids = true,
symbolic x coords = {2003/04,2004/05,2005/06,2006/07,2007/08,2008/09,2009/10,2010/11,2011/12,2012/13,2013/14,
2014/15,2015/16,2016/17,2017/18,2018/19,2019/20,2020/21,2021/22,2022/23,2023/24,2024/25,2025/26},
xtick = {2003/04,2004/05,2005/06,2006/07,2007/08,2008/09,2009/10,2010/11,2011/12,2012/13,2013/14,
2014/15,2015/16,2016/17,2017/18,2018/19,2019/20,2020/21,2021/22,2022/23,2023/24,2024/25,2025/26},
xticklabel style = {rotate=60},
xlabel = {Season},
xlabel style = {align=center, font=\small},
enlarge x limits = 0.02,
ymin = 0,
%ytick distance = 1,
scaled y ticks = false,
ylabel = {Relative frequency (\%)},
ylabel style = {align=center, font=\small},
yticklabel style = {/pgf/number format/fixed,/pgf/number format/precision=5},
ytick style = {draw = none},
legend style = {font=\small,at={(0.1,-0.35)},anchor=north west,legend columns=2},
legend entries = {Old group stage$\qquad$, New incomplete round-robin league}
]
% Group stage
\addplot [blue, only marks, mark = star, thick] coordinates{
(2003/04,23.9583333333333)
(2004/05,21.875)
(2005/06,26.0416666666667)
(2006/07,26.0416666666667)
(2007/08,18.75)
(2008/09,29.1666666666667)
(2009/10,26.0416666666667)
(2010/11,17.7083333333333)
(2011/12,29.1666666666667)
(2012/13,20.8333333333333)
(2013/14,18.75)
(2014/15,21.875)
(2015/16,17.7083333333333)
(2016/17,31.25)
(2017/18,20.8333333333333)
(2018/19,25)
(2019/20,22.9166666666667)
(2020/21,20.8333333333333)
(2021/22,18.75)
(2022/23,19.7916666666667)
(2023/24,20.8333333333333)
%(2024/25,12.5)
%(2025/26,17.36111111111111)
};
% Group stage
\addplot [red, only marks, mark = o, thick] coordinates{
(2024/25,12.5)
(2025/26,17.36111111111111)
};
\end{axis}
\end{tikzpicture}

\captionsetup{justification=centering}
\caption{Proportion of draws in the first phase of the \\ UEFA Champions League between 2003/04 and 2025/26}
\label{Fig1}

\end{figure}

%\end{document}

The motivation behind the current study comes from a surprising empirical observation. Figure~\ref{Fig1} presents the proportion of draws in the UEFA Champions League (UCL) -- the most prestigious European club football competition -- since 2003/04. After the format change, the relative frequency of draws declined from 23\% (the average over the 21 seasons in the old format) to 15\% (the average in the new format). These numbers are robust to the exclusion of matches between teams drawn from the same seeding pot introduced by the new format.
Our paper aims to investigate whether this decrease can be explained by a rational reaction from the clubs, induced by stronger incentives for offensive play after the reform. In particular, we consider the following research question: \textit{How has the relative value of a win and a loss compared to a draw shifted as a result of using the innovative incomplete round-robin tournament format instead of the classical group stage?}

Although this setting seems to provide an ideal natural experiment, data availability precludes any reliable statistical test for several years to come. Therefore, a simulation-based approach, calibrated based on the participating teams and results from previous UCL seasons, will be adopted. Our framework consists of three steps: we (1) randomly draw realistic sets of clubs and match schedules; (2) generate hypothetical match outcomes; and (3) compute the expected payoff from attacking. In step (3), the marginal effects of scoring and conceding a goal on the qualification probabilities for the two critical ranking thresholds -- implied by the qualification rules -- are compared in each of the two formats. Comparing potential gains and losses allows quantifying the (relative) incentive to choose the risky strategy of attacking in any single match.

We find that the new UCL design has substantially increased the difference between a win and a draw relative to that between a draw and a loss, which encourages offensive play and can explain the decline in the frequency of draws on the field. Even though roughly half of the increase is driven by the modified ranking thresholds, the remaining half is directly attributable to the format change. Thus, replacing the group stage with an incomplete round-robin league offers a promising way for contest designers to strengthen incentives to attack without changing the proportion of qualifying teams.

Our roadmap is as follows. As promised, Section~\ref{Sec2} gives a literature overview on the impact of the three-points rule and incomplete round-robin tournaments. The simulation methodology is detailed in Section~\ref{Sec3}, and the results are presented in Section~\ref{Sec4}. Section~\ref{Sec5} concludes, discusses the limitations of the study, and outlines directions for future research.

\section{Related literature} \label{Sec2}
%\label{sec:literature}
%\input{files_DW/02_Literature_DW}

This paper is connected to two almost orthogonal research fields.
First, tournament organisers often seek to influence the behaviour of competitors by changing the reward system. Therefore, it would be important to understand the impact of three points for a win. That rule was introduced in English football in 1981 and was adopted in other sports, such as ice hockey, volleyball, and water polo, in the 1990s. A comparison of German football league and cup competitions suggests that the reform achieved its main aim: the percentage of matches ending in a draw decreased significantly \citep{DilgerGeyer2009}. \citet{Moschini2010} develops a game-theoretic model and tests its implications on a large empirical dataset. A significant increase is found in both the expected number of goals and a reduction in the frequency of draws. According to the regression discontinuity method of \citet{AlfanoCicatielloGaetaGalloRotondo2021}, the rule also worked in Italian football as expected.

However, the positive results have been challenged several times.
Clearly, conditional on the match being tied, increasing the points for a win implies a more offensive strategy toward the end of the match. On the other hand, a more defensive strategy could be optimal at the beginning of the match to break the tie only later, which might be especially appealing to the weaker teams \citep{BrocasCarrillo2004}. If the asymmetry between the opponents is sufficiently high, an increase in the reward for victory induces the underdog to play more defensively; this hypothesis is supported by historical data from the Portuguese football league \citep{GuedesMachado2002}. Analogously, the three-points rule results in a more defensive match if the home advantage is sufficiently strong and the teams' levels of offensiveness are strategic substitutes \citep{DewenterNamini2013}. According to the theoretical model of \citet{Jost2021b}, the effect of winning points on the number of draws depends crucially on team heterogeneity.

In addition, teams leading by one goal should become more conservative, since conceding a goal costs two points rather than one, which is consistent with the empirical results of \citet{GaricanoPalacios-Huerta2005}: these teams significantly increase their number of defenders, which leads to a lower probability of scoring an additional goal and fewer opponent attempts on goal.
\citet{RiedlHeuerStrauss2015} use the principle of loss aversion from prospect theory to show that the three-points rule failed to reverse the preference to avoid defeats over achieving victories. Indeed, football players weigh potential losses more than potential gains and adjust their effort accordingly \citep{SchneemannDeutscher2017}.
Based on a regression discontinuity design applied to the German Bundesliga, \citet{HonParinduri2016} do not find any evidence for the three-points rule making matches more decisive, increasing the number of goals, or decreasing goal difference.
Using a quasi-experimental design based on a synthetic control method, the reform is shown to have improved competitive balance in England but to have had no significant effect on the number of goals scored in a match \citep{Sharma2025}. Last but not least, \citet{StraussHagemannLoffing2009} argue that large fluctuations in the percentage of draws in the German football league preclude any identification of the effect of the three-points rule on their proportion.

The controversy around the consequences of increasing the points for a win has two important messages for us. First, empirical data may be influenced by several factors whose complex interactions are difficult (or even impossible) to disentangle by a statistical estimation. Second, a reliable econometric analysis calls for a large amount of data that will be available only years or even decades later, since incomplete round-robin tournaments have emerged only in recent years. This calls for a novel approach to quantify incentives, which will be presented in Section~\ref{Sec35}.

The second set of related papers analyse the incomplete round-robin format.
\citet{LiVanBulckGoossens2025} propose organising a single, incomplete round-robin contest involving all teams rather than splitting them into round-robin groups or leagues. Compared to the classical solution to the multi-league timetabling problem in sports competitions with a huge number of teams, the flexibility of the incomplete round-robin is effective in reducing capacity violations and travel distance. Since reducing travel distance is especially relevant to youth sports, \citet{DevriesereGoossens2025} study the redesign of Belgian youth field hockey competitions as incomplete round-robin tournaments. Total travel time can decrease by up to 25\%, and a 20\% gain can be ensured even if the disadvantage of the worst-off team is strictly limited.

The new incomplete round-robin league phase of the UEFA Champions League has also been studied. \citet{Gyimesi2024} examines its effect on short-, mid-, and long-term competitive balance. The 2024/25 reform is found to increase match uncertainty and reduce the proportion of matches in which at least one team is indifferent to the outcome. \citet{DevriesereGoossensSpieksma2026} compare the old and new UCL formats by estimating the proportion of three unattractive categories of matches: asymmetric (exactly one of the two teams is indifferent), stakeless (the outcome has no impact on the prizes of both teams), and collusive (both teams have something to play for but a particular result is beneficial for both of them). The incomplete round-robin format implies more competitive and collusive matches. According to \citet{CsatoDevriesereGoossensGyimesiLambersSpieksma2026}, the official ranking can be debated in the 2024/25 UEFA Champions League due to its disregard for the strength of the schedule and its unfair favouring of teams playing against a weaker set of opponents.

\citet{GuyonBenSalemBuchholtzerTanre2026} investigate the draw of the UCL league phase. They prove that the scheduling constraint cannot be ignored in the first stage of the draw, which determines the set of matches to be played. The minimum number of breaks (the sum of consecutive home or away matches for each team) is shown to be four, and a schedule is constructed to minimise breaks, while guaranteeing various fairness and attractiveness criteria. Four different draw procedures are compared in terms of fairness and matchup probabilities. The authors also provide an interactive draw simulator at \url{https://julienguyon.github.io/UEFA-league-phase-draw/}.
\citet{CsatoGyimesiGoossensDevriesereLambersSpieksma2026} investigate the effects of the draw on the variance of qualifying probabilities for each team. The new UCL format is less sensitive to the draw, but mainly due to the inaccurate seeding policy \citep{Csato2024c}.
\citet{deWerraUrrutiaAssuncao2026} derive some bounds on the minimum number of breaks in incomplete round-robin tournaments, and \citet{DevriesereGoossensSpieksma2027} study the problem of finding break-free timetables.
\citet{WesselSpieksma2026} consider the complexity of deciding whether a fair draw exists. According to their simulations, the bandwidth (the difference between the strongest and the weakest opponent sets) of draws by randomly generated opponent sets is an order of magnitude worse than the minimum possible bandwidth.

Finally, although the aims are completely different, our methodological approach adopts elements from two working papers, \citet{CsatoGyimesi2026b} and \citet{WinkelmannDeutscherMichels2026}.
\citet{CsatoGyimesi2026b} also compute the increase and decrease in the probability of obtaining a prize from a win and from a loss, respectively, using a goalless draw as a reference. Here we consider not only whether their ratio is below or above the natural threshold of one, but focus on how it changes, and the analysis is not restricted to the last matchday.
\citet{WinkelmannDeutscherMichels2026} propose a simulation model to predict qualification thresholds in the new design of the UEFA Champions League and the UEFA Europa League, the second-tier European club football competition. Its main feature is accounting for the reduced probability of draws. While their goal is to investigate the implications of a lower draw frequency, our paper focuses on exploring the reasons for this observation.
% by considering the incentives of teams, using the same modelling framework.

\section{Methodology} \label{Sec3}
%\label{sec:method}
%\input{files_DW/03_Method_DW}

This section presents our framework used to compare the incentives to attack in the two fundamentally different UCL formats. First, Section~\ref{Sec31} presents the details of both tournament designs. We then introduce the general simulation setup in Section~\ref{Sec32} and explain how the sets of participating clubs and match schedules are constructed in Section~\ref{Sec33}. Section~\ref{Sec34} describes the prediction model for match outcomes, and Section~\ref{Sec35} shows the quantification of incentives.

\subsection{Tournament designs} \label{Sec31}
%\label{sec:method_design}

In the 21 seasons between 2003/04 and 2023/24, the UCL was organised in the same format and started with a group stage. For the group draw, 32 teams were seeded into four pots of eight teams each. The seeding was primarily based on ranking the clubs by their UEFA club coefficients, the official measure of team strength, except for Pot~1. Pot~1 contained (a) the titleholder plus the seven strongest teams until the 2014/15 season; (b) the titleholder and the champions of the seven highest-ranked associations between 2015/16 and 2017/18; and (c) the titleholder, the UEFA Europa League titleholder, as well as the champions of the six highest-ranked associations between 2018/19 and 2023/24 \citep{Csato2021a}. Vacancies were filled by the eighth-strongest team before 2014/15 and by the champion(s) of the next highest-ranked association(s) after 2015/16 \citep{Csato2020a}. Teams from the same country were not allowed to play in the same group. UEFA also formed some pairs of teams from the same national association to ensure that their matches were played on different days \citep{Guyon2021a}. The groups were organised in a double round-robin format, i.e.\ each team played against all the others once at home and once away. The top two teams qualified for the Round of 16, where the group winners meet the runners-up in two-legged ties, with the first leg hosted by the runners-up. The third-placed teams were transferred to the secondary UEFA competition (UEFA Cup until 2009; UEFA Europa League thereafter), and the fourth-placed teams were eliminated.

Starting from the 2024/25 season, 36 clubs compete in an incomplete round-robin league phase with eight rounds. Prior to the draw, the teams are seeded into four pots of nine teams each based on their UEFA club coefficients, except for the guaranteed slot of the titleholder in Pot~1. Every team plays against two opponents from each pot, one at home and the other away. Teams are not allowed to play against a team from the same association, and can play against at most two opponents from the same country. Finally, the matches should be played over eight matchdays, that is, the chromatic number of the match graph -- where the nodes represent the clubs and the edges represent the matches -- needs to be eight \citep{GuyonBenSalemBuchholtzerTanre2026}. All teams are ranked in a single table. The first eight teams directly qualify for the Round of 16. In the knockout phase play-offs, the seeded teams ranked 9--16 play two-legged ties against the unseeded teams ranked 17--24, with the unseeded teams hosting the first leg. The eight winners of the play-offs also qualify for the Round of 16. The last 12 teams in the league phase are eliminated.

%Consequently, four major changes have been implemented in 2024:
%\begin{itemize}
%\item
%The eight groups have been replaced by a single incomplete round-robin league;
%\item 
%The number of clubs has been increased from 32 to 36;
%\item 
%Each club plays two extra matches against clubs drawn from the same pot;
%\item 
%The prizes have become more exclusive as the proportion of clubs qualifying directly for the Round of 16 (eliminated) has decreased (increased).
%\end{itemize}
%Section~\ref{Sec45} will measure the impact of the most important individual changes sequentially.

\subsection{Simulation setup} \label{Sec32}

Quantifying incentives for offensive play starts with simulating the two tournament designs.
In the old format, we consider 1,000 different groups (4,000 teams), corresponding to 125 seasons (years) of the UCL. Analogously, 125 seasons are considered in the new format with 4,500 teams in total.

At first, for each season, the Elo ratings of the participating clubs, a proxy for their strength, are generated randomly. Details on the generation of clubs and match schedules are provided in Section~\ref{Sec33}.
For any given season with a specific set of teams and a schedule, we perform 10,000 simulation runs. In each run, all match outcomes are generated by the underlying Elo-based prediction model presented in Section~\ref{Sec34}. We then calculate the final standings and rank the teams according to the official ranking criteria.

Hence, in the old format, the clubs are ranked based on their points, followed by head-to-head results (points collected in matches among tied teams, goal difference in matches among tied teams, and goals scored in matches among tied teams). Head-to-head results are used recursively: if more than two teams are tied and applying all head-to-head criteria yields a subset of them tied, the same criteria are applied again to that subset only. The next tie-breaking criteria are overall goal difference and goals scored.
In the new format, teams are ranked in a single table based on their number of points. The tie-breaking rules are goal difference, goals scored, away goals scored, wins, away wins (in the 2024/25 season, the order of Real Madrid and Bayern Munich was decided by the number of away wins).
Any remaining ties are resolved randomly in both formats since implementing further criteria would be unreasonable.

This enables us to determine, for each team and simulation run, whether they finished in the top two (three) in the old format and in the top eight (24) in the new format, the ranking thresholds justified in Section~\ref{Sec35}. A natural binary coding is used for both prizes, where 1 corresponds to achievement and 0 to non-achievement.

\subsection{Generating team strengths and the set of matches} \label{Sec33}

%We simulate 125-125 UCL seasons in both the old and new formats.
To follow the structural design of the UCL, we have to take into account that the clubs are drawn from four seeding pots. Since the allocation rule for Pot~1 varied over time (see Section~\ref{Sec31}), we always assign the teams to seeding pots strictly based on their club coefficients to ensure a fair comparison.

Team strengths are drawn from a set determined by historical data.
The Elo ratings of all clubs participating in the UCL between 2015/16 and 2025/26, separated by the seeding pot, give $9 \times 8 + 2 \times 9 = 90$ values for each pot. In the old format, we randomly sample one of these 90 values for each group as the proxy for team strength from the corresponding seeding pot in each simulated season. Accordingly, in the new format, we draw nine independent values (with replacement) per seeding pot and season.
Due to the allocation of the four extra slots, it is a realistic assumption that one team is added to each of the four pots without changing the distribution of Elo values  \citep{CsatoGyimesiGoossensDevriesereLambersSpieksma2026}.

The old format is a double round-robin, where the set of matches is given. However, this is more complicated in the new format: the draw requires, for example, solving successive integer programming problems \citep{DevriesereGoossensSpieksma2026, GuyonBenSalemBuchholtzerTanre2026}.
Hence, our implementation does not exactly replicate the official draw mechanism, but closely mimics it. We guarantee that the primary constraints are satisfied, i.e.\ each club plays against two different opponents from every seeding pot, with one match played at home and one away. Association restrictions are ignored. The draw is performed in sequential blocks. First, all eight opponents from Pots~1--4 are determined for teams in Pot~1. Next, the remaining six opponents from Pots~2--4 are determined for teams in Pot~2, and so on. If a deadlock occurs in the draw, which is possible \citep{Csato2025c}, we discard the corresponding block and repeat the draw for that block only. One simulation run requires, on average, about 1.21 redraws.
The order of matches does not affect our simulation results.
% As the match prediction model relies on fixed Elo ratings

\subsection{Generating match results} \label{Sec34}

The simulation of a tournament usually relies on a statistical model fitted to historical team strengths and match outcomes, allowing hypothetical results to be generated for given team strengths. In other words, the identities of the clubs are ignored, and they are represented solely by their ratings.

The classical choice for the distribution of the number of goals in a match is the independent Poisson model \citep{Maher1982}. Let $\lambda_{jk}$ be the expected number of goals scored by team $j$ against team $k$. The probability of team $j$ scoring $m$ goals against team $k$ is
\begin{equation}
P_{jk}(m) = \frac{\lambda_{jk}^{m} \exp{\left( -\lambda_{jk} \right)}}{m!}.
\end{equation}

Following \citet{DagaevRudyak2019}, the single parameter of the Poisson distribution is assumed to depend on the difference between the strengths of the opposing teams and the field of the match. Therefore, in a match played by the home team $j$ against the away team $k$:
\begin{equation}
\log \left( \lambda_{jk} \right) = \alpha + \beta \cdot \left(R_j -  R_k \right) + \gamma,
\end{equation}
\begin{equation}
\log \left( \lambda_{kj} \right) = \alpha + \beta \cdot \left(R_k -  R_j\right),
\end{equation}
where $R_j$ and $R_k$ are the Elo ratings, while $\alpha$, $\beta$, and $\gamma$ are parameters to be estimated.

However, the number of goals scored by the two teams is not necessarily independent. For example, if one team scores, its opponent may take more risks to score, leading to a positive correlation \citep{KarlisNtzoufras2003}. While \citet{CsatoMolontayPinter2024} show that relatively simple models could outperform more complex ones in the old UCL format, \citet{WinkelmannDeutscherMichels2026} demonstrate the aptitude of the Dixon--Coles model \citep{DixonColes1997} to the new UCL format. Therefore, we adopt this model.

Let $X$ and $Y$ denote the goals scored by teams $j$ and $k$, respectively. The probability of the scoreline $(x,y)$ is
\begin{equation}
P(X=x, Y=y) = \tau_{\rho}(x,y) \cdot \frac{\lambda_{jk}^x e^{-\lambda_{jk}}}{x!} \cdot \frac{\lambda_{kj}^y e^{-\lambda_{kj}}}{y!}.
\end{equation}
Parameter $\tau_{\rho}(x,y)$ corrects the probability of four low-scoring outcomes (0-0, 0-1, 1-0, 1-1):
\begin{equation}
\tau_{\rho}(x,y) = 
\begin{cases} 
1 - \lambda_{jk} \lambda_{kj} \rho & \text{if } x=y=0,\\
1 + \lambda_{jk} \rho & \text{if } x=0, y=1,\\
1 + \lambda_{kj} \rho & \text{if } x=1, y=0,\\
1 - \rho & \text{if } x=y=1,\\
1 & \text{otherwise}.
\end{cases}
\end{equation}
$\rho$ shifts the probability mass from the outcomes of 0–0 and 1–1 towards the outcomes of 1–0 and 0–1 ($\rho > 0$), or vice versa ($\rho < 0$). While the Dixon--Coles model was originally applied to competitions with a higher proportion of draws, it can also be used in the opposite case \citep{MichelsOttingKarlis2025}, which is relevant to the new UCL format, see Figure~\ref{Fig1}.
For $\rho = 0$, the Dixon--Coles model collapses to the independent model of \citet{Maher1982} .

\subsection{Incentives for offensive play and the definition of prizes} \label{Sec35}

%For each match and simulation run, match outcomes are randomly drawn based on the Elo ratings of the hypothetical clubs (Section~\ref{Sec33}) and the underlying prediction model (Section~\ref{Sec34}). Simulating the 12 matches per group (old format) and 144 per season (new format) allows computing the final standings (Section~\ref{Sec32}).

Any football match starts with the result of 0-0. When a team chooses between the risky strategy of playing more offensively and the conservative strategy of focusing on defence, the potential gain from scoring a goal should be compared with the potential loss from conceding one. The incentive to attack is stronger if the former is higher relative to the latter. Thus, the main challenge in modelling resides in quantifying these gains and losses.

We define them through the implied changes in the probability of obtaining a given prize compared to the probability under the goalless draw of 0-0. For a match between teams $j$ and $k$, denote this probability by
(a) $p_{jk}^w$ if team $j$ wins by 1-0;
(b) $p_{jk}^d$ if teams $j$ and $k$ play a draw of 0-0; and
(c) $p_{jk}^{\ell}$ if team $j$ loses by 0-1.
The incentive of team $j$ to attack against team $k$ is the \emph{expected payoff}
\begin{equation} \label{eq:incentives}
\mathcal{I}_{jk} = \frac{p_{jk}^w - p_{jk}^d}{p_{jk}^d - p_{jk}^{\ell}},
\end{equation}
which is the increase in the probability of obtaining the prize if the match is won, divided by the decrease in the probability of obtaining the prize if the match is lost.
While retaining the simulated results of all other matches, the outcome of an individual match is set to either 1-0, 0-0, or 0-1. Then we calculate the final standings and determine the prizes for both teams, which enables us to compute both $\mathcal{I}_{jk}$ and $\mathcal{I}_{kj}$ according to Equation~\eqref{eq:incentives}.

In the group stage of the UCL, the two main prizes were qualifying for the Round of 16 (top two) and avoiding elimination from European competitions (top three). Analogously, the two main prizes in the new league phase of the UCL are direct qualification for the Round of 16 (being ranked in the top eight) and avoiding elimination (being ranked in the top 24 but outside the top eight). This definition is consistent with recent studies on match classification \citep{CsatoGyimesi2026b, DevriesereGoossensSpieksma2026}.
%We will focus on the associated probabilities in the following. With a slight abuse of notation, they are denoted by $\mathcal{I}^2$, $\mathcal{I}^3$ (group stage), and $\mathcal{I}^8$, $\mathcal{I}^{24}$ (league phase).

For each of the 125 seasons, the number of cases -- out of the 10,000 simulation runs -- is counted in which a team, given a certain result in a specific match, obtains the prize considered. We further aggregate across all clubs from the same seeding pot (eight in the old format and nine in the new format). Finally, this value is divided by the number of clubs and the number of simulation runs. Averaging over all seasons provides an estimate of the probability of winning the specific prize, given one of three outcomes (1-0, 0-0, 0-1) in a match. In the last step, we calculate the incentive for offensive play using Equation~\ref{eq:incentives}.

\section{Results} \label{Sec4}

In the following, we introduce the data used for the simulations (Section~\ref{Sec41}) and present the estimated parameters of the match prediction model (Section~\ref{Sec42}). Afterwards, Section~\ref{Sec43} discusses the resulting probabilities for obtaining the two prizes in the old and new formats of the UCL, before Section~\ref{Sec44} explicitly investigates incentives for offensive play. Finally, Section~\ref{Sec45} assigns the changes in incentives to some components of the 2024 UCL reform. 

\subsection{Team strengths} \label{Sec41}

As outlined in Section~\ref{Sec33}, our prediction model relies on Elo ratings from the 11 UCL seasons between 2015/16 and 2025/26. The data is retrieved from Football Club Elo Ratings (\url{www.clubelo.com}), a widely used measure of team strength in the literature \citep{BoskerGurtler2024, Csato2022b, Csato2024c, CsatoPetroczy2025a, CsatoPetroczy2026, Rewilak2026, YildirimBilman2025a}.

\begin{table}[t!]
    \centering
    \caption{Summary statistics of Elo ratings in recent UCL seasons}
    \label{Table1}
    \rowcolors{1}{}{gray!20}
    \begin{threeparttable}
    \begin{tabularx}{\textwidth}{Ll CCCC C} \toprule
         Format & Statistic & Pot 1 & Pot 2 & Pot 3 & Pot 4 & All \\ \bottomrule
        & Min & 1681 & 1672 & 1443 & 1334 & 1334 \\
        & Max & 2094 & 2065 & 1901 & 1974 & 2094 \\
        Old & Mean & 1936 & 1816 & 1722 & 1640 & 1779 \\
        & Median & 1953 & 1811 & 1735 & 1653 & 1763 \\
        & Standard deviation & 92 & 76 & 84 & 110 & 143 \\ \hline
        & Min & 1861 & 1572 & 1553 & 1446 & 1446 \\
        & Max & 2060 & 1950 & 1835 & 1792 & 2060 \\
        New & Mean & 1927 & 1807 & 1688 & 1708 & 1782 \\
        & Median & 1905 & 1836 & 1686 & 1769 & 1791 \\
        & Standard deviation & 65 & 115 & 106 & 115 & 137 \\ \bottomrule
    \end{tabularx}
\begin{tablenotes} \footnotesize
\item
\textit{Notes:}
Old format: UCL group stage from 2015/16 to 2023/24.
New format: UCL league phase in 2024/25 and 2025/26.
\item
Football Club Elo Ratings are collected from 1 September of the corresponding season.
\end{tablenotes}
\end{threeparttable}
\end{table}

Table~\ref{Table1} provides summary statistics on Elo ratings in the old and new formats, respectively, separated and aggregated by seeding pot. On average, Elo values decrease from Pot~1 to Pot~4, as they strongly correlate with UEFA club coefficients. The standard deviation is the highest for Pot~4. There are only minor differences in the Elo ratings of participating clubs between the old and new formats, which is unsurprising, as the entry rules have changed only slightly.

\subsection{Estimation of the match prediction model} \label{Sec42}

The Dixon--Coles model presented in Section~\ref{Sec34} is fitted to the $9 \times 8 \times 12 = 864$ group stage matches played between 2015/16 and 2023/24 in the old format, as well as to the $2 \times 144 = 288$ league phase matches played in the first two seasons in the new format.

\begin{table}[t!]
\centering 
  \caption{Estimated coefficients of the Dixon--Coles model in the UCL} 
  \label{Table2}
\begin{threeparttable}  
\begin{tabularx}{0.7\textwidth}{LCC} \toprule 
 & \multicolumn{2}{c}{\textit{Dependent variable: Log goals}} \\ 
 & Old format & New format \\ \midrule
\multirow{2}{*}{Intercept ($\alpha$)} & 0.208*** & 0.286*** \\ 
  & (0.032) & (0.051) \\
\multirow{2}{*}{Elo difference ($\beta$)} & 0.0017*** & 0.0018*** \\ 
  & (0.00009) & (0.0002)\\ 
\multirow{2}{*}{Home field ($\gamma$)} & 0.268*** & 0.289*** \\ 
  & (0.041) & (0.065)\\ 
\multirow{2}{*}{Parameter $\rho$} & $-$0.162*** & 0.036 \\ 
  & (0.050) & (0.071)\\ \midrule 
Observations & 864 & 288 \\ \bottomrule
\end{tabularx}
\begin{tablenotes} \footnotesize
\item
\textit{Notes:}
Old format: UCL group stage from 2015/16 to 2023/24.
New format: UCL league phase in 2024/25 and 2025/26.
%\item
%Adjustment parameter $\rho$ shifts probability mass from 0-0 and 1-1 outcomes towards 1-0 and 0-1 outcomes (vice versa) if $\rho > 0$ ($\rho < 0$).
\item
Standard errors are in parenthesis.
Significance: * $p<0.10$, ** $p<0.05$, *** $p<0.01$.
\end{tablenotes}
\end{threeparttable}
\end{table}

Table~\ref{Table2} reports the estimated coefficients. Naturally, a larger difference in the strengths of the opposing teams is expected to increase (decrease) the number of goals scored by the stronger (weaker) team. Home teams score significantly more goals on average; home advantage is equivalent to about 150 Elo points ($\approx \gamma / \beta$). For a balanced match with equal Elo ratings, home teams are expected to score 1.61 (1.78) goals in the old (new) format, whereas away teams are expected to score 1.23 (1.33) goals. This aligns with the empirical observation reported by \citet{WinkelmannDeutscherMichels2026} that, on average, more goals are scored in the new format. 

Parameter $\rho$ is of particular interest as it is responsible for the shift of probability between low-scoring draws and wins. 
%Specifically, positive values correspond to shifting probability mass from 0-0 and 1-1 outcomes towards 1-0 and 0-1 outcomes (and vice versa for negative values). 
In the old format, its value is significantly negative; more low-scoring draws occur than the independent Poisson model of \citet{Maher1982} would imply. In the new format, the effect is positive but insignificant. The insignificance may be driven by the small magnitude of the effect, the limited sample size, or a combination of both.
A standard two-sided Wald test rejects the null hypothesis: the difference of $\rho$ between the old and new formats is statistically significant (p-value $\approx 0.023$). Given the example with equal Elo ratings for the teams, the estimated parameters suggest that a 0-0 (1-1) draw is 32.1\% (16.2\%) more probable in the old UCL format than in the independent Poisson model of \citet{Maher1982}. However, the chances of these outcomes are lower by 8.5\% (3.6\%) in the new UCL format.
%The sensitivity analyses of \citet{WinkelmannDeutscherMichels2026} demonstrate the impact of different values of $\rho$ on the expected frequency of draws in the new UCL format.

When simulating the new incomplete round-robin format with parameter estimates from the group stage (left column of Table~\ref{Table2}), the model gives 35.3 draws per season (24.5\%) on average, which is clearly above the empirical observations in the 2024/25 and 2025/26 seasons.
To conclude, the statistical tests show that teams are less likely to settle for a draw and more likely to push for a win in the new UCL format, which strengthens the assumption that Figure~\ref{Fig1} documents a regime change.

\subsection{Prize achievement prospects} \label{Sec43}

Using the estimated coefficients from Table~\ref{Table2} and the simulation framework outlined in Section~\ref{Sec32}, we now compute the probabilities of achieving the first and second prizes in the old and new UCL formats.
According to Section~\ref{Sec35}, the three potential outcomes of 1-0, 0-0, and 0-1 are considered for each match from the perspective of both teams. The (36 out of 144) matches played by clubs from drawn the same seeding pot in the new format are excluded from the analysis, as these matches did not take place in the old format. In total, this gives $3 \times 2 \times 12 = 72$ probabilities for each prize and format, which are reported in Tables~\ref{Table_A1} (first prize) and Tables~\ref{Table_A2} (second prize) in Appendix~\ref{Appendix_A}.

\begin{figure}[t!]
\centering

\begin{subfigure}{0.49\textwidth}
\centering
\caption{Old group stage format}
\label{Fig2a}

\begin{tikzpicture}
%\selectcolormodel{gray}
\begin{axis}[
%name = axis1,
title = {First team plays at home},
title style = {font=\small},
width = \textwidth, 
height = 0.6\textwidth,
xmajorgrids,
ymajorgrids,
%xbar stacked,
%bar width = 10pt,
scaled x ticks = false,
symbolic x coords = {1-2,1-3,1-4,2-1,2-3,2-4,3-1,3-2,3-4,4-1,4-2,4-3},
xticklabel style = {rotate=45, anchor=north east},
xtick = data,
enlarge x limits = 0.05,
ylabel = {Probability (\%)},
ylabel style = {align=center, font=\small},
ymin = 0,
ymax = 100,
ybar stacked,
bar width = 10pt,
]
% Loss
\addplot [forget plot, transparent] coordinates{
(1-2,72.02105)
(1-3,67.30672)
(1-4,67.45594)
(2-1,50.31621)
(2-3,35.5642)
(2-4,37.15778)
(3-1,25.69489)
(3-2,17.98507)
(3-4,18.06751)
(4-1,11.32972)
(4-2,7.73766)
(4-3,7.38942)
};
% Draw - Loss
\addplot [red, thick, pattern = vertical lines, pattern color = red] coordinates{
(1-2,10.00453)
(1-3,12.91422)
(1-4,11.3900200000001)
(2-1,10.35819)
(2-3,18.75189)
(2-4,15.09058)
(3-1,9.90479000000001)
(3-2,14.24482)
(3-4,10.17895)
(4-1,5.91419999999999)
(4-2,7.51926)
(4-3,6.51039)
};
% Win - Draw
\addplot [ForestGreen, thick, pattern = crosshatch dots, pattern color = ForestGreen] coordinates{
(1-2,10.0800700000001)
(1-3,11.15986)
(1-4,10.5293799999999)
(2-1,16.26187)
(2-3,20.7303800000001)
(2-4,17.79346)
(3-1,19.0114)
(3-2,22.4414)
(3-4,16.2914)
(4-1,14.28378)
(4-2,15.59326)
(4-3,13.0064)
};
\end{axis}
\end{tikzpicture}

\vspace{0.25cm}
\begin{tikzpicture}
%\selectcolormodel{gray}
\begin{axis}[
%name = axis1,
title = {First team plays away},
title style = {font=\small},
width = \textwidth, 
height = 0.6\textwidth,
xmajorgrids,
ymajorgrids,
%xbar stacked,
%bar width = 10pt,
scaled x ticks = false,
xlabel = {Matches between teams \\ drawn from different pots},
xlabel style = {align=center, font=\small},
symbolic x coords = {1-2,1-3,1-4,2-1,2-3,2-4,3-1,3-2,3-4,4-1,4-2,4-3},
xticklabel style = {rotate=45, anchor=north east},
xtick = data,
enlarge x limits = 0.05,
ylabel = {Probability (\%)},
ylabel style = {align=center, font=\small},
ymin = 0,
ymax = 100,
ybar stacked,
bar width = 10pt,
legend style = {font=\small,at={(0.25,-0.65)},anchor=north west,legend columns=3},
legend entries = {Loss$\qquad$, Win},
]
% Loss
\addplot [forget plot, transparent] coordinates{
(1-2,76.6218)
(1-3,73.24252)
(1-4,72.63552)
(2-1,54.57829)
(2-3,43.358)
(2-4,43.63915)
(3-1,29.4299)
(3-2,23.37825)
(3-4,22.2995)
(4-1,13.42712)
(4-2,10.49755)
(4-3,9.94007)
};
% Draw - Loss
\addplot [red, thick, pattern = vertical lines, pattern color = red] coordinates{
(1-2,9.40139000000002)
(1-3,11.71274)
(1-4,10.08133)
(2-1,11.03575)
(2-3,19.35181)
(2-4,14.84117)
(3-1,11.63855)
(3-2,16.9877)
(3-4,10.8834)
(4-1,7.39309999999999)
(4-2,9.654)
(4-3,7.89421)
};
% Win - Draw
\addplot [ForestGreen, thick, pattern = crosshatch dots, pattern color = ForestGreen] coordinates{
(1-2,8.21071000000001)
(1-3,8.75067999999998)
(1-4,8.68921999999998)
(2-1,15.36425)
(2-3,17.89546)
(2-4,15.91193)
(3-1,19.44892)
(3-2,22.01831)
(3-4,16.11543)
(4-1,15.66578)
(4-2,16.89065)
(4-3,13.92342)
};
\end{axis}
\end{tikzpicture}
\end{subfigure}
\begin{subfigure}{0.49\textwidth}
\centering
\caption{New incomplete round-robin format}
\label{Fig2b}

\begin{tikzpicture}
%\selectcolormodel{gray}
\begin{axis}[
%name = axis1,
title = {First team plays at home},
title style = {font=\small},
width = \textwidth, 
height = 0.6\textwidth,
xmajorgrids,
ymajorgrids,
%xbar stacked,
%bar width = 10pt,
scaled x ticks = false,
symbolic x coords = {1-2,1-3,1-4,2-1,2-3,2-4,3-1,3-2,3-4,4-1,4-2,4-3},
xticklabel style = {rotate=45, anchor=north east},
xtick = data,
enlarge x limits = 0.05,
ylabel = {Probability (\%)},
ylabel style = {align=center, font=\small},
ymin = 0,
ymax = 100,
ybar stacked,
bar width = 10pt,
]
% Loss
\addplot [forget plot, transparent] coordinates{
(1-2,32.2272355555556)
(1-3,30.2368088888889)
(1-4,28.7429777777778)
(2-1,12.9688533333333)
(2-3,9.79358222222223)
(2-4,8.93256888888889)
(3-1,4.63644444444444)
(3-2,3.71386666666667)
(3-4,2.85037333333333)
(4-1,2.07664)
(4-2,1.67544888888889)
(4-3,1.46394666666667)
};
% Draw - Loss
\addplot [red, thick, pattern = vertical lines, pattern color = red] coordinates{
(1-2,9.94980444444442)
(1-3,9.32180444444444)
(1-4,8.73641777777778)
(2-1,6.58215111111111)
(2-3,5.08555555555555)
(2-4,4.56408)
(3-1,3.28843555555556)
(3-2,2.69082666666667)
(3-4,1.98610666666667)
(4-1,1.42828444444444)
(4-2,1.14852444444444)
(4-3,0.956426666666667)
};
% Win - Draw
\addplot [ForestGreen, thick, pattern = crosshatch dots, pattern color = ForestGreen] coordinates{
(1-2,18.5881511111111)
(1-3,18.1361777777778)
(1-4,17.6669066666667)
(2-1,15.9498044444444)
(2-3,13.3858222222222)
(2-4,12.41072)
(3-1,9.80897777777779)
(3-2,8.39799111111112)
(3-4,6.62957333333333)
(4-1,4.73585777777778)
(4-2,3.91478222222222)
(4-3,3.37746666666667)
};
\end{axis}
\end{tikzpicture}

\vspace{0.25cm}
\begin{tikzpicture}
%\selectcolormodel{gray}
\begin{axis}[
%name = axis1,
title = {First team plays away},
title style = {font=\small},
width = \textwidth, 
height = 0.6\textwidth,
xmajorgrids,
ymajorgrids,
%xbar stacked,
%bar width = 10pt,
scaled x ticks = false,
xlabel = {Matches between teams \\ drawn from different pots},
xlabel style = {align=center, font=\small},
symbolic x coords = {1-2,1-3,1-4,2-1,2-3,2-4,3-1,3-2,3-4,4-1,4-2,4-3},
xticklabel style = {rotate=45, anchor=north east},
xtick = data,
enlarge x limits = 0.05,
ylabel = {Probability (\%)},
ylabel style = {align=center, font=\small},
ymin = 0,
ymax = 100,
ybar stacked,
bar width = 10pt,
legend style = {font=\small,at={(0.25,-0.65)},anchor=north west,legend columns=3},
legend entries = {Loss$\qquad$, Win},
]
% Loss
\addplot [forget plot, transparent] coordinates{
(1-2,37.8424088888889)
(1-3,34.8459377777778)
(1-4,32.21304)
(2-1,16.7781955555556)
(2-3,12.3582222222222)
(2-4,10.7904711111111)
(3-1,6.49671111111111)
(3-2,5.19427555555556)
(3-4,3.64681777777778)
(4-1,2.90698666666667)
(4-2,2.30109333333333)
(4-3,1.94815111111111)
};
% Draw - Loss
\addplot [red, thick, pattern = vertical lines, pattern color = red] coordinates{
(1-2,10.3501511111111)
(1-3,9.82689777777777)
(1-4,9.2160977777778)
(2-1,7.48077333333333)
(2-3,5.97185777777778)
(2-4,5.25650666666666)
(3-1,4.03344)
(3-2,3.4168)
(3-4,2.43879111111111)
(4-1,1.79144)
(4-2,1.49555555555555)
(4-3,1.24204444444444)
};
% Win - Draw
\addplot [ForestGreen, thick, pattern = crosshatch dots, pattern color = ForestGreen] coordinates{
(1-2,18.2224711111111)
(1-3,18.3159555555555)
(1-4,18.1528888888889)
(2-1,16.8940355555556)
(2-3,15.0004622222222)
(2-4,13.9153955555556)
(3-1,11.2289333333333)
(3-2,10.0467466666667)
(3-4,7.92601777777779)
(4-1,5.65046222222222)
(4-2,4.89367111111112)
(4-3,4.27518222222222)
};
\end{axis}
\end{tikzpicture}
\end{subfigure}

\captionsetup{justification=centerfirst}
\caption{Probability of obtaining the first prize in the old and new UCL formats \\ \vspace{0.2cm}
\footnotesize{\emph{Notes:} Matches between teams from different seeding pots are distinguished. \\
In each pair $a$-$b$, the first team drawn from the higher-ranked pot $a$ plays at home (top panels) or away (bottom panels). \\
A draw of 0-0 is the reference category.}}
\label{Fig2}

\end{figure}

%\end{document}                                                                                                                                                                

Figure~\ref{Fig2} compares the probabilities for the first prize (advancing directly to the Round of 16) in the old (Figure~\ref{Fig2a}) and new (Figure~\ref{Fig2b}) formats. The initial scoreline of 0-0 is set as the baseline at the boundary of the two bars. The top green (bottom red) bars indicate a positive (negative) shift in the probability after winning (losing) the specific match.
Clubs from the top seeding pots have a higher probability of achieving this prize, as they are typically stronger. However, while the probabilities for teams from Pot~1 are 67\%--92\% in the old format, they are clearly lower in the new format, ranging from 30\% to 66\%. A similar pattern can be seen for the other seeding pots. Again, this is not surprising, since the first prize was awarded to two out of four teams in the group stage, whereas it is awarded to only eight out of 36 teams in the league phase. The decline is remarkable especially for Pot~4 teams, from 7--37\% in the old to at most 10\% in the new format.

Even though we find the highest gain from a single match when a team from Pot~2 beats a team from Pot~3 and vice versa (up to 40 percentage points) in the old format, single matches are the most important with respect to achieving the first prize for Pot~1 teams in the new format, with a difference of up to 29 percentage points. This is mainly because teams from lower pots have at most a 43\% chance of obtaining the first prize, even when one win is already secured. Note that the number of matches per team is higher in the new format, which immediately reduces the impact of an extra win compared with the previous group stage -- and justifies the focus on relative changes in Equation~\eqref{eq:incentives}.

\begin{figure}[t!]
\centering

\begin{subfigure}{0.49\textwidth}
\centering
\caption{Old group stage format}
\label{Fig3a}

\begin{tikzpicture}
%\selectcolormodel{gray}
\begin{axis}[
%name = axis1,
title = {First team plays at home},
title style = {font=\small},
width = \textwidth, 
height = 0.6\textwidth,
xmajorgrids,
ymajorgrids,
%xbar stacked,
%bar width = 10pt,
scaled x ticks = false,
symbolic x coords = {1-2,1-3,1-4,2-1,2-3,2-4,3-1,3-2,3-4,4-1,4-2,4-3},
xticklabel style = {rotate=45, anchor=north east},
xtick = data,
enlarge x limits = 0.05,
ylabel = {Probability (\%)},
ylabel style = {align=center, font=\small},
ymin = 0,
ymax = 100,
ybar stacked,
bar width = 10pt,
]
% Loss
\addplot [forget plot, transparent] coordinates{
(1-2,92.73825)
(1-3,90.84817)
(1-4,87.73315)
(2-1,83.53599)
(2-3,76.26086)
(2-4,69.22224)
(3-1,66.84709)
(3-2,62.40623)
(3-4,46.35349)
(4-1,36.47996)
(4-2,31.30875)
(4-3,24.1716)
};
% Draw - Loss
\addplot [red, thick, pattern = vertical lines, pattern color = red] coordinates{
(1-2,2.90441000000004)
(1-3,4.20080999999995)
(1-4,6.55438000000008)
(2-1,4.77215999999999)
(2-3,9.16613999999991)
(2-4,14.2834500000001)
(3-1,7.25017999999999)
(3-2,9.51418999999998)
(3-4,20.40008)
(4-1,8.60831000000001)
(4-2,11.19449)
(4-3,15.63915)
};
% Win - Draw
\addplot [ForestGreen, thick, pattern = crosshatch dots, pattern color = ForestGreen] coordinates{
(1-2,2.75435999999998)
(1-3,3.31709)
(1-4,4.04371999999992)
(2-1,6.10532999999998)
(2-3,8.83943000000005)
(2-4,10.7728599999999)
(3-1,10.81378)
(3-2,13.0059)
(3-4,18.81008)
(4-1,15.94355)
(4-2,19.54302)
(4-3,23.4755)
};
\end{axis}
\end{tikzpicture}

\vspace{0.25cm}
\begin{tikzpicture}
%\selectcolormodel{gray}
\begin{axis}[
%name = axis1,
title = {First team plays away},
title style = {font=\small},
width = \textwidth, 
height = 0.6\textwidth,
xmajorgrids,
ymajorgrids,
%xbar stacked,
%bar width = 10pt,
scaled x ticks = false,
xlabel = {Matches between teams \\ drawn from different pots},
xlabel style = {align=center, font=\small},
symbolic x coords = {1-2,1-3,1-4,2-1,2-3,2-4,3-1,3-2,3-4,4-1,4-2,4-3},
xticklabel style = {rotate=45, anchor=north east},
xtick = data,
enlarge x limits = 0.05,
ylabel = {Probability (\%)},
ylabel style = {align=center, font=\small},
ymin = 0,
ymax = 100,
ybar stacked,
bar width = 10pt,
legend style = {font=\small,at={(0.25,-0.65)},anchor=north west,legend columns=3},
legend entries = {Loss$\qquad$, Win},
]
% Loss
\addplot [forget plot, transparent] coordinates{
(1-2,94.13139)
(1-3,92.88377)
(1-4,90.88564)
(2-1,85.51895)
(2-3,80.3955)
(2-4,75.67311)
(3-1,69.54245)
(3-2,66.29194)
(3-4,54.9575)
(4-1,39.32336)
(4-2,35.444)
(4-3,30.0211)
};
% Draw - Loss
\addplot [red, thick, pattern = vertical lines, pattern color = red] coordinates{
(1-2,2.57031999999999)
(1-3,3.61622)
(1-4,5.36751000000001)
(2-1,4.51864000000006)
(2-3,8.68891999999996)
(2-4,12.9404)
(3-1,7.29654000000003)
(3-2,9.95741000000004)
(3-4,20.42575)
(4-1,9.36912999999999)
(4-2,13.23204)
(4-3,18.79316)
};
% Win - Draw
\addplot [ForestGreen, thick, pattern = crosshatch dots, pattern color = ForestGreen] coordinates{
(1-2,2.17334999999999)
(1-3,2.44054999999997)
(1-4,2.76043)
(2-1,5.45459999999999)
(2-3,7.02931)
(2-4,7.82376000000001)
(3-1,10.28646)
(3-2,11.95525)
(3-4,14.91735)
(4-1,16.29328)
(4-2,19.68071)
(4-3,22.56654)
};
\end{axis}
\end{tikzpicture}
\end{subfigure}
\begin{subfigure}{0.49\textwidth}
\centering
\caption{New incomplete round-robin format}
\label{Fig3b}

\begin{tikzpicture}
%\selectcolormodel{gray}
\begin{axis}[
%name = axis1,
title = {First team plays at home},
title style = {font=\small},
width = \textwidth, 
height = 0.6\textwidth,
xmajorgrids,
ymajorgrids,
%xbar stacked,
%bar width = 10pt,
scaled x ticks = false,
symbolic x coords = {1-2,1-3,1-4,2-1,2-3,2-4,3-1,3-2,3-4,4-1,4-2,4-3},
xticklabel style = {rotate=45, anchor=north east},
xtick = data,
enlarge x limits = 0.05,
ylabel = {Probability (\%)},
ylabel style = {align=center, font=\small},
ymin = 0,
ymax = 100,
ybar stacked,
bar width = 10pt,
]
% Loss
\addplot [forget plot, transparent] coordinates{
(1-2,88.0394666666666)
(1-3,86.1062755555555)
(1-4,84.2567733333333)
(2-1,72.1516622222223)
(2-3,64.4543644444445)
(2-4,61.1977422222222)
(3-1,50.6845422222222)
(3-2,46.0858577777778)
(3-4,38.8777955555555)
(4-1,29.0939555555555)
(4-2,25.58704)
(4-3,22.8270133333333)
};
% Draw - Loss
\addplot [red, thick, pattern = vertical lines, pattern color = red] coordinates{
(1-2,4.10902222222227)
(1-3,4.67770666666671)
(1-4,5.0856177777778)
(2-1,7.55166222222218)
(2-3,9.04797333333335)
(2-4,9.33432)
(3-1,9.40423111111114)
(3-2,9.8993688888889)
(3-4,9.87852444444446)
(4-1,8.01453333333335)
(4-2,7.96772444444444)
(4-3,7.71206222222221)
};
% Win - Draw
\addplot [ForestGreen, thick, pattern = crosshatch dots, pattern color = ForestGreen] coordinates{
(1-2,4.86522666666666)
(1-3,5.59825777777779)
(1-4,6.2171911111111)
(2-1,11.0373066666667)
(2-3,13.6177333333333)
(2-4,14.4052266666666)
(3-1,16.7897066666666)
(3-2,17.9606666666666)
(3-4,18.7314044444444)
(4-1,17.8102133333333)
(4-2,18.0799377777778)
(4-3,17.8017866666667)
};
\end{axis}
\end{tikzpicture}

\vspace{0.25cm}
\begin{tikzpicture}
%\selectcolormodel{gray}
\begin{axis}[
%name = axis1,
title = {First team plays away},
title style = {font=\small},
width = \textwidth, 
height = 0.6\textwidth,
xmajorgrids,
ymajorgrids,
%xbar stacked,
%bar width = 10pt,
scaled x ticks = false,
xlabel = {Matches between teams \\ drawn from different pots},
xlabel style = {align=center, font=\small},
symbolic x coords = {1-2,1-3,1-4,2-1,2-3,2-4,3-1,3-2,3-4,4-1,4-2,4-3},
xticklabel style = {rotate=45, anchor=north east},
xtick = data,
enlarge x limits = 0.05,
ylabel = {Probability (\%)},
ylabel style = {align=center, font=\small},
ymin = 0,
ymax = 100,
ybar stacked,
bar width = 10pt,
legend style = {font=\small,at={(0.25,-0.65)},anchor=north west,legend columns=3},
legend entries = {Loss$\qquad$, Win},
]
% Loss
\addplot [forget plot, transparent] coordinates{
(1-2,90.6825066666667)
(1-3,89.1935733333333)
(1-4,87.3733866666667)
(2-1,76.0447733333333)
(2-3,70.1314666666666)
(2-4,66.7122311111111)
(3-1,55.2585066666667)
(3-2,51.8152711111111)
(3-4,44.64528)
(4-1,32.7228888888889)
(4-2,29.9959822222222)
(4-3,27.4594222222222)
};
% Draw - Loss
\addplot [red, thick, pattern = vertical lines, pattern color = red] coordinates{
(1-2,3.36673777777778)
(1-3,3.88664000000005)
(1-4,4.42565333333333)
(2-1,6.81015999999998)
(2-3,8.3406311111111)
(2-4,8.87371555555558)
(3-1,9.05713777777779)
(3-2,9.69161777777776)
(3-4,10.1668266666667)
(4-1,8.1841688888889)
(4-2,8.41092444444443)
(4-3,8.39780444444444)
};
% Win - Draw
\addplot [ForestGreen, thick, pattern = crosshatch dots, pattern color = ForestGreen] coordinates{
(1-2,3.79335111111108)
(1-3,4.36639999999998)
(1-4,5.0954044444445)
(2-1,9.57672000000005)
(2-3,11.7335644444445)
(2-4,12.8834311111111)
(3-1,15.5054311111111)
(3-2,16.5968444444445)
(3-4,18.2512533333333)
(4-1,17.5150755555555)
(4-2,18.0422044444445)
(4-3,18.2326577777777)
};
\end{axis}
\end{tikzpicture}
\end{subfigure}

\captionsetup{justification=centerfirst}
\caption{Probability of obtaining the second prize in the old and new UCL formats \\ \vspace{0.2cm}
\footnotesize{\emph{Notes:} Matches between teams from different seeding pots are distinguished. \\
In each pair $a$-$b$, the first team drawn from the higher-ranked pot $a$ plays at home (top panels) or away (bottom panels). \\
A draw of 0-0 is the reference category.}}
\label{Fig3}

\end{figure}

%\end{document}                                                                                                                                                                

Figure~\ref{Fig3} provides the same comparison for the second prize, which is transfer to the UEFA Europa League in the old format and qualification for the knockout phase play-offs in the new format. For teams drawn from Pots~1 and 2, there are only minor differences in these probabilities between the two formats. However, teams from Pot~4 again have considerably lower probabilities to achieve the prize, with a reduction of up to 21 percentage points. The reason is that, in the new format, only the top 24 clubs (67\%) receive the second prize instead of three out of four in the groups (75\%). Consequently, the strongest effect of a single match is observed in the group stage between Pot~3 and Pot~4 teams, but this does not hold for the incomplete round-robin format.

Appendix~\ref{Appendix_B} verifies that the conclusions do not change when the reference scoreline is set to 1-1 or 2-2 instead of 0-0.

\subsection{Comparison of the incentives for offensive play} \label{Sec44}

\begin{figure}[t!]
\centering

\begin{subfigure}{\textwidth}
\centering
\caption{Home team}
\label{Fig4a}

\begin{tikzpicture}
\begin{axis}[
width = \textwidth, 
height = 0.4\textwidth,
%title = Unweighted rounds,
%title style = {font=\small},
xmajorgrids = true,
ymajorgrids = true,
symbolic x coords = {1-2,1-3,1-4,2-1,2-3,2-4,3-1,3-2,3-4,4-1,4-2,4-3},
xtick = data,
xlabel = {Matches between teams drawn from different pots},
xlabel style = {align=center, font=\small},
enlarge x limits = 0.04,
ymin = 0,
%ytick distance = 1,
scaled y ticks = false,
ylabel = {Expected payoff \\ for offensive play},
ylabel style = {align=center, font=\small},
yticklabel style = {/pgf/number format/fixed,/pgf/number format/precision=5},
ytick style = {draw = none},
]
% Top 2 group
\addplot [blue, mark = square, mark options = {solid, thick}, only marks, thick] coordinates{
(1-2,1.00755057958746)
(1-3,0.864152848565383)
(1-4,0.924439114242108)
(2-1,1.56995285855927)
(2-3,1.10550883137647)
(2-4,1.17911041192585)
(3-1,1.91941474781393)
(3-2,1.57540776226025)
(3-4,1.60049906915743)
(4-1,2.41516688647661)
(4-2,2.0737758768815)
(4-3,1.99779122295285)
};
% Top 8 league
\addplot [ForestGreen, mark = triangle, mark options = {solid, thick}, only marks, thick, x filter/.code={\pgfmathparse{\pgfmathresult+0.1}\pgfmathresult}] coordinates{
(1-2,1.86819260769391)
(1-3,1.94556514094076)
(1-4,2.02221403738324)
(2-1,2.42319025728839)
(2-3,2.63212584662443)
(2-4,2.71921613994496)
(3-1,2.98287061189515)
(3-2,3.12097067237495)
(3-4,3.3379744626002)
(4-1,3.31576654510151)
(4-2,3.4085319134116)
(4-3,3.53133887246975)
};
% Top 3 group
\addplot [red, mark = star, mark options = {solid, thick}, only marks, thick, x filter/.code={\pgfmathparse{\pgfmathresult-0.2}\pgfmathresult}] coordinates{
(1-2,0.948337183799789)
(1-3,0.789631047345639)
(1-4,0.61694927666688)
(2-1,1.27936406155703)
(2-3,0.964356861230587)
(2-4,0.75421974382939)
(3-1,1.4915188312566)
(3-2,1.36700023859099)
(3-4,0.922059129179886)
(4-1,1.85211150620737)
(4-2,1.74577135715875)
(4-3,1.50107262862751)
};
% Top 24 league
\addplot [brown, mark = pentagon, mark options = {solid, thick}, only marks, thick, x filter/.code={\pgfmathparse{\pgfmathresult+0.2}\pgfmathresult}] coordinates{
(1-2,1.18403513136403)
(1-3,1.19679539071377)
(1-4,1.22250459684128)
(2-1,1.46157314004158)
(2-3,1.50505895979651)
(2-4,1.54325399886297)
(3-1,1.78533539513183)
(3-2,1.81432441484485)
(3-4,1.89617432742991)
(4-1,2.22223959806351)
(4-2,2.26914696960739)
(4-3,2.30830433594934)
};
\end{axis}
\end{tikzpicture}
\end{subfigure}

\vspace{0.25cm}
\begin{subfigure}{\textwidth}
\centering
\caption{Away team}
\label{Fig4b}

\begin{tikzpicture}
\begin{axis}[
width = \textwidth, 
height = 0.4\textwidth,
%title = Unweighted rounds,
%title style = {font=\small},
xmajorgrids = true,
ymajorgrids = true,
symbolic x coords = {1-2,1-3,1-4,2-1,2-3,2-4,3-1,3-2,3-4,4-1,4-2,4-3},
xtick = data,
xlabel = {Matches between teams drawn from different pots},
xlabel style = {align=center, font=\small},
enlarge x limits = 0.04,
ymin = 0,
%ytick distance = 1,
scaled y ticks = false,
ylabel = {Expected payoff \\ for offensive play},
ylabel style = {align=center, font=\small},
yticklabel style = {/pgf/number format/fixed,/pgf/number format/precision=5},
ytick style = {draw = none},
legend style = {font=\small,at={(0,-0.25)},anchor=north west,legend columns=2},
legend entries = {{$\mathcal{I}_{jk}$ group stage, first prize}$\quad \qquad$, {$\mathcal{I}_{jk}$ incomplete round-robin, first prize}$\quad$, {$\mathcal{I}_{jk}$ group stage, second prize}$\qquad$, {$\mathcal{I}_{jk}$ incomplete round-robin, second prize}}
]
% Top 2 group
\addplot [blue, mark = square, mark options = {solid, thick}, only marks, thick] coordinates{
(1-2,0.873350642830474)
(1-3,0.747107850084608)
(1-4,0.86191206914167)
(2-1,1.39222526787939)
(2-3,0.924743473607898)
(2-4,1.07214795059958)
(3-1,1.67107758268856)
(3-2,1.29613249586465)
(3-4,1.48073488064391)
(4-1,2.11897309653596)
(4-2,1.74960120157447)
(4-3,1.76375090097679)
};
% Top 8 league
\addplot [ForestGreen, mark = triangle, mark options = {solid, thick}, only marks, thick, x filter/.code={\pgfmathparse{\pgfmathresult+0.1}\pgfmathresult}] coordinates{
(1-2,1.76059952318463)
(1-3,1.86385937553707)
(1-4,1.96969360857475)
(2-1,2.25832742188271)
(2-3,2.51185858411453)
(2-4,2.6472705996547)
(3-1,2.78395943247782)
(3-2,2.94039647233279)
(3-4,3.24997813124171)
(4-1,3.15414539265742)
(4-2,3.27214264487371)
(4-3,3.44205252987905)
};
% Top 3 group
\addplot [red, mark = star, mark options = {solid, thick}, only marks, thick, x filter/.code={\pgfmathparse{\pgfmathresult-0.2}\pgfmathresult}] coordinates{
(1-2,0.845556195337542)
(1-3,0.674889802058495)
(1-4,0.514285022291528)
(2-1,1.20713311969971)
(2-3,0.808996975458404)
(2-4,0.604599548700197)
(3-1,1.40977230303678)
(3-2,1.20063851945435)
(3-4,0.730320796054002)
(4-1,1.73903873678773)
(4-2,1.48735266822047)
(4-3,1.20078475360184)
};
% Top 24 league
\addplot [brown, mark = pentagon, mark options = {solid, thick}, only marks, thick, x filter/.code={\pgfmathparse{\pgfmathresult+0.2}\pgfmathresult}] coordinates{
(1-2,1.12671415521255)
(1-3,1.12343823971346)
(1-4,1.1513338394733)
(2-1,1.40624008833861)
(2-3,1.40679575539715)
(2-4,1.45186433241543)
(3-1,1.71195707645682)
(3-2,1.71249473772067)
(3-4,1.79517699393583)
(4-1,2.14011658280104)
(4-2,2.14509172726687)
(4-3,2.17112197579685)
};
\end{axis}
\end{tikzpicture}
\end{subfigure}

\captionsetup{justification=centerfirst}
\caption{Incentives for offensive play in the old and new UCL formats by seeding pots \\ \vspace{0.2cm}
\footnotesize{\emph{Notes:} Matches between teams from different seeding pots are distinguished. \\
In each pair $a$-$b$, the first team drawn from the higher-ranked pot $a$ plays at home (top panel) or away (bottom panel). \\
Values are based on Equation~\ref{eq:incentives}.}}
\label{Fig4}

\end{figure}

%\end{document}

Based on Tables~\ref{Table_A1} and~\ref{Table_A2} in Appendix~\ref{Appendix_A}, the expected payoff of offensive play can be easily computed according to Equation~\eqref{eq:incentives}. %Specifically, we consider $\mathcal{I}^2$ and $\mathcal{I}^3$ in the old group stage format, and $\mathcal{I}^8$ and $\mathcal{I}^{24}$ in the new incomplete round-robin format as the first and second prizes, respectively, for the 12 different types of matches. 
Figure~\ref{Fig4} illustrates the results for home (Figure~\ref{Fig4a}) and away (Figure~\ref{Fig4b}) teams. Our incentive measure consistently shows that it is more beneficial to attack in the league phase if the aim is to obtain the first prize. The difference is non-negligible; the incentives are 37\% to 172\% stronger, depending on the match type. While some large values are driven by the extremely low probability of achieving the first prize for Pot~4 teams, we still observe a consistent increase, even for the strongest Pot~1 teams. Their incentives for offensive play were below one in almost all matches in the previous group stage, suggesting that they might have been satisfied with a draw in individual matches. Hence, they could have safely substituted and rested some of their top players in certain matches, which is unfair and detrimental to attractiveness.
However, in the new format, this ratio ranges from 1.76 to 2.02, encouraging them to play more offensively.
For the second prize, the incentives in the league phase again robustly exceed the incentives in the group stage for all matches. Particularly large differences, exceeding 100\%, are observed in matches between the weakest teams from Pot~3 and Pot~4.

On average, the weaker team has a stronger incentive to attack, which is intuitive under the assumption of equal risk of scoring and conceding a goal. Generally, the strongest teams drawn from Pot~1 are the least encouraged to score, as they can obtain the prize with a relatively high probability that does not depend much on the outcome of a particular match. In contrast, the weakest teams drawn from Pot~4 could lose little by attacking, while their potential gain is substantial, especially for the top eight positions in the league phase, which is the most challenging achievement. The same pattern holds for both home and away teams, with slightly but consistently higher incentives for home teams, because losing at home is more detrimental than losing away due to home advantage. 

Although analysing matches by seeding pots is natural, there is relatively strong variation in Elo ratings even within a seeding pot (see Table~\ref{Table1}). Therefore, Appendix~\ref{Appendix_C} examines the incentives for offensive play in both the old and new UCL formats by clustering matches based on Elo ratings. In the old format, we find a $\cup$-shaped effect as a function of opponent strength. Similar to Figure~\ref{Fig4}, the incentives are usually weaker for strong teams.
In the new format, the incentives for offensive play are again consistently stronger than in the old format, but they vary less with the strength of the opponent. Thus, the conclusions drawn from Figure~\ref{Fig4} would not change if Elo ratings were used instead of seeding pots. 

To summarise, the new incomplete round-robin league undeniably creates stronger incentives to attack than the previous group stage for both prizes and (almost) all types of matches. According to our numerical results, the average increase reaches 100\% relative to (direct) qualification for the Round of 16 (first prize), and 59\% relative to avoiding elimination from European competitions (second prize), calculated over the 12 pairs of seeding pots and the two types (home, away) of teams.

\subsection{Decomposing the changes in the incentives for offensive play} \label{Sec45}

In the baseline analysis, we compared the incentives for offensive play in the old and new UCL formats used.
However, this comparison is not able to disentangle the effect of the following changes presented in the two sets of simulations:
\begin{itemize}
\item
Instead of eight groups, an incomplete round-robin tournament with a single ranking table is used to decide which teams receive the prizes;
\item 
The number of teams has increased from 32 to 36, and teams play eight matches against opponents from all four seeding pots, rather than six matches against teams from the three pots that do not contain the given team;
\item 
The first prize was awarded to the top two out of four teams (50\%) and the second prize to the top three teams (75\%) in each group, while eight teams out of 36 (22.2\%) receive the first prize and 24 (66.7\%) receive the second prize now;
\item 
The prediction model fitted to the new format has different parameters than the model estimated based on the old format (see Table~\ref{Table2}).
\end{itemize}

\begin{table}[t!]
    \centering
    \caption{Overview of alternative UCL formats}
    \label{Table3}
    \begin{threeparttable}
    \rowcolors{1}{}{gray!20}
    \begin{tabularx}{\textwidth}{llCCcc} \toprule
        ID & Competition type & Clubs & Same pot & Prizes in \% & Training sample \\ \bottomrule
        1 & Group stage & 32 & No & 50 (75) & 2015/16--2023/24 \\
        2 & Incomplete round-robin & 32 & No & 50 (75) & 2015/16--2023/24 \\
        3 & Incomplete round-robin & 36 & Yes & 50 (75) & 2015/16--2023/24 \\
        4 & Incomplete round-robin & 36 & Yes & 22.2 (66.7) & 2015/16--2023/24 \\
        5 & Incomplete round-robin & 36 & Yes & 22.2 (66.7) & 2024/25--2025/26 \\ \toprule
    \end{tabularx}
\begin{tablenotes} \footnotesize
\item
\textit{Notes:}
Column \textit{ID} is the identifier of the tournament format. Column \textit{Clubs} is the number of teams. Column \textit{Same pot} indicates whether teams drawn from the same pot play against each other. Column \textit{Prizes in \%} gives the proportion of teams that obtain the first (second) prize. Column \textit{Training sample} is the set of seasons used to estimate the parameters of the Dixon--Coles model, see Table~\ref{Table2}.
\end{tablenotes}
\end{threeparttable}
\end{table}

In order to decompose the aggregated effects of these changes presented in Figure~\ref{Fig4}, we consider three intermediate variants, Formats 2--4, see Table~\ref{Table3}. Format~1 (top row) corresponds to the UCL format used until the 2023/24 season, whereas Format~5 (bottom row) is the design from the 2024/25 season onward.

\begin{figure}[t!]
\centering

\begin{subfigure}{\textwidth}
\centering
\caption{Home team}
\label{Fig5a}

\begin{tikzpicture}
%\selectcolormodel{gray}
\begin{axis}[
name = axis1,
title = {First prize},
title style = {font=\small},
width = 0.47\textwidth, 
height = 0.5\textwidth,
xmajorgrids,
ymajorgrids,
%xbar stacked,
%bar width = 10pt,
scaled x ticks = false,
xlabel = {Expected payoff for offensive play},
xlabel style = {align=center, font=\small},
xticklabel style = {/pgf/number format/fixed,/pgf/number format/precision=5},
xmin = 0.5,
xmax = 4,
ylabel = {Matches between teams \\ drawn from different pots},
ylabel style = {align=center, font=\small},
symbolic y coords = {1-2,1-3,1-4,2-1,2-3,2-4,3-1,3-2,3-4,4-1,4-2,4-3},
ytick = data,
y dir = reverse,
enlarge y limits = 0.05,
]
% Format 1
\addplot [red, mark = +, mark options = {solid, thick}, only marks] coordinates{
(1.00755057958746,1-2)
(0.864152848565383,1-3)
(0.924439114242108,1-4)
(1.56995285855927,2-1)
(1.10550883137647,2-3)
(1.17911041192585,2-4)
(1.91941474781393,3-1)
(1.57540776226025,3-2)
(1.60049906915743,3-4)
(2.41516688647661,4-1)
(2.0737758768815,4-2)
(1.99779122295285,4-3)
};
% Format 2
\addplot [blue, mark = x, mark options = {solid, thick}, only marks, y filter/.code={\pgfmathparse{\pgfmathresult-0.2}\pgfmathresult}] coordinates{
(1.18218626313734,1-2)
(1.19659591111796,1-3)
(1.27004499790708,1-4)
(1.65561416479071,2-1)
(1.73500036547491,2-3)
(1.83388807876252,2-4)
(2.24654002416143,3-1)
(2.33614149936627,3-2)
(2.51186740953532,3-4)
(3.07718717336404,4-1)
(3.1709868025183,4-2)
(3.2020314493151,4-3)
};
% Format 3
\addplot [brown, mark = otimes, mark options = {solid, thick}, only marks, y filter/.code={\pgfmathparse{\pgfmathresult+0.2}\pgfmathresult}] coordinates{
(1.31758261492321,1-2)
(1.34108746282489,1-3)
(1.39883995590251,1-4)
(1.6759970687511,2-1)
(1.74304406111074,2-3)
(1.80995437270225,2-4)
(2.03996924859063,3-1)
(2.08140041901966,3-2)
(2.20759486693272,3-4)
(2.4852770095748,4-1)
(2.53571428571429,4-2)
(2.59619716123692,4-3)
};
% Format 4
\addplot [ForestGreen, mark = Mercedes star, mark options = {solid, thick}, only marks, y filter/.code={\pgfmathparse{\pgfmathresult-0.2}\pgfmathresult}] coordinates{
(1.86829873975225,1-2)
(1.94648703858082,1-3)
(2.02326375790483,1-4)
(2.43201770968323,2-1)
(2.63014824978083,2-3)
(2.718269693375,2-4)
(3.011126505542,3-1)
(3.1352314760911,3-2)
(3.3507938213673,3-4)
(3.38078060188862,4-1)
(3.48157522719624,4-2)
(3.56889321433093,4-3)
};
% Format 5
\addplot [black, mark = diamond, mark options = {solid, thick}, only marks, y filter/.code={\pgfmathparse{\pgfmathresult+0.2}\pgfmathresult}] coordinates{
(1.86819260769391,1-2)
(1.94556514094076,1-3)
(2.02221403738324,1-4)
(2.42319025728839,2-1)
(2.63212584662443,2-3)
(2.71921613994496,2-4)
(2.98287061189515,3-1)
(3.12097067237495,3-2)
(3.3379744626002,3-4)
(3.31576654510151,4-1)
(3.4085319134116,4-2)
(3.53133887246975,4-3)
};
\end{axis}

\begin{axis}[
at = {(axis1.south east)},
xshift = 0.125\textwidth,
title = {Second prize},
title style = {font=\small},
width = 0.47\textwidth, 
height = 0.5\textwidth,
xmajorgrids,
ymajorgrids,
%xbar stacked,
%bar width = 10pt,
scaled x ticks = false,
xlabel = {Expected payoff for offensive play},
xlabel style = {align=center, font=\small},
xticklabel style = {/pgf/number format/fixed,/pgf/number format/precision=5},
xmin = 0.4,
xmax = 2.6,
ylabel = {Matches between teams \\ drawn from different pots},
ylabel style = {align=center, font=\small},
symbolic y coords = {1-2,1-3,1-4,2-1,2-3,2-4,3-1,3-2,3-4,4-1,4-2,4-3},
ytick = data,
y dir = reverse,
enlarge y limits = 0.05,
]
% Format 1
\addplot [red, mark = +, mark options = {solid, thick}, only marks] coordinates{
(0.948337183799789,1-2)
(0.789631047345639,1-3)
(0.61694927666688,1-4)
(1.27936406155703,2-1)
(0.964356861230587,2-3)
(0.75421974382939,2-4)
(1.4915188312566,3-1)
(1.36700023859099,3-2)
(0.922059129179886,3-4)
(1.85211150620737,4-1)
(1.74577135715875,4-2)
(1.50107262862751,4-3)
};
% Format 2
\addplot [blue, mark = x, mark options = {solid, thick}, only marks, y filter/.code={\pgfmathparse{\pgfmathresult-0.2}\pgfmathresult}] coordinates{
(0.876488696300089,1-2)
(0.865107838166832,1-3)
(0.862885876671464,1-4)
(1.12114523817627,2-1)
(1.14429409460653,2-3)
(1.16047489087875,2-4)
(1.48023753609567,3-1)
(1.53332509707629,3-2)
(1.55293694811518,3-4)
(2.22062455085908,4-1)
(2.31490343578591,4-2)
(2.33480619440089,4-3)
};
% Format 3
\addplot [brown, mark = otimes, mark options = {solid, thick}, only marks, y filter/.code={\pgfmathparse{\pgfmathresult+0.2}\pgfmathresult}] coordinates{
(1.00738335515414,1-2)
(1.01033917351106,1-3)
(1.03055971168022,1-4)
(1.21560483955519,2-1)
(1.24766858678445,2-3)
(1.27599178749475,2-4)
(1.49001689773033,3-1)
(1.51424868160082,3-2)
(1.56598630442838,3-4)
(1.88089440545669,4-1)
(1.91007522224755,4-2)
(1.92904580152672,4-3)
};
% Format 4
\addplot [ForestGreen, mark = Mercedes star, mark options = {solid, thick}, only marks, y filter/.code={\pgfmathparse{\pgfmathresult-0.2}\pgfmathresult}] coordinates{
(1.14311126586783,1-2)
(1.15513126491648,1-3)
(1.17909210424562,1-4)
(1.41241727687112,2-1)
(1.45562078372101,2-3)
(1.49194955123524,2-4)
(1.73318845322801,3-1)
(1.76006017427737,3-2)
(1.83508569306187,3-4)
(2.17560449711887,4-1)
(2.21883255716156,4-2)
(2.24911532722234,4-3)
};
% Format 5
\addplot [black, mark = diamond, mark options = {solid, thick}, only marks, y filter/.code={\pgfmathparse{\pgfmathresult+0.2}\pgfmathresult}] coordinates{
(1.18403513136403,1-2)
(1.19679539071377,1-3)
(1.22250459684128,1-4)
(1.46157314004158,2-1)
(1.50505895979651,2-3)
(1.54325399886297,2-4)
(1.78533539513183,3-1)
(1.81432441484485,3-2)
(1.89617432742991,3-4)
(2.22223959806351,4-1)
(2.26914696960739,4-2)
(2.30830433594934,4-3)
};
\end{axis}
\end{tikzpicture}
\end{subfigure}

\begin{subfigure}{\textwidth}
\centering
\caption{Away team}
\label{Fig5b}

\begin{tikzpicture}
%\selectcolormodel{gray}
\begin{axis}[
name = axis1,
title = {First prize},
title style = {font=\small},
width = 0.47\textwidth, 
height = 0.5\textwidth,
xmajorgrids,
ymajorgrids,
%xbar stacked,
%bar width = 10pt,
scaled x ticks = false,
xlabel = {Expected payoff for offensive play},
xlabel style = {align=center, font=\small},
xticklabel style = {/pgf/number format/fixed,/pgf/number format/precision=5},
xmin = 0.5,
xmax = 4,
ylabel = {Matches between teams \\ drawn from different pots},
ylabel style = {align=center, font=\small},
symbolic y coords = {1-2,1-3,1-4,2-1,2-3,2-4,3-1,3-2,3-4,4-1,4-2,4-3},
ytick = data,
y dir = reverse,
enlarge y limits = 0.05,
legend style = {font=\small,at={(0,-0.2)},anchor=north west,legend columns=5},
legend entries = {Format 1$\qquad$, Format 2$\qquad$, Format 3$\qquad$, Format 4$\qquad$, Format 5},
]
ó% Format 1
\addplot [red, mark = +, mark options = {solid, thick}, only marks] coordinates{
(0.873350642830474,1-2)
(0.747107850084608,1-3)
(0.86191206914167,1-4)
(1.39222526787939,2-1)
(0.924743473607898,2-3)
(1.07214795059958,2-4)
(1.67107758268856,3-1)
(1.29613249586465,3-2)
(1.48073488064391,3-4)
(2.11897309653596,4-1)
(1.74960120157447,4-2)
(1.76375090097679,4-3)
};
% Format 2
\addplot [blue, mark = x, mark options = {solid, thick}, only marks, y filter/.code={\pgfmathparse{\pgfmathresult-0.2}\pgfmathresult}] coordinates{
(1.10157290380787,1-2)
(1.10779521951635,1-3)
(1.19686016332216,1-4)
(1.53656217378273,2-1)
(1.58490901769792,2-3)
(1.72669387392701,2-4)
(2.05156101878998,3-1)
(2.09965538620238,3-2)
(2.3762258930953,3-4)
(2.80804296627806,4-1)
(2.90156815064761,4-2)
(3.03070873244194,4-3)
};
% Format 3
\addplot [brown, mark = otimes, mark options = {solid, thick}, only marks, y filter/.code={\pgfmathparse{\pgfmathresult+0.2}\pgfmathresult}] coordinates{
(1.25017949617035,1-2)
(1.27079505382586,1-3)
(1.33173034383433,1-4)
(1.59954823001023,2-1)
(1.64732128240555,2-3)
(1.72455140289082,2-4)
(1.93871514964585,3-1)
(1.95877993324156,3-2)
(2.11412768121949,3-4)
(2.36109638110905,4-1)
(2.39870910385555,4-2)
(2.4637833326403,4-3)
};
% Format 4
\addplot [ForestGreen, mark = Mercedes star, mark options = {solid, thick}, only marks, y filter/.code={\pgfmathparse{\pgfmathresult-0.2}\pgfmathresult}] coordinates{
(1.77290523416071,1-2)
(1.87032115475268,1-3)
(1.9635322474947,1-4)
(2.27583282943069,2-1)
(2.53027297629168,2-3)
(2.64496056864585,2-4)
(2.7967665098676,3-1)
(2.9669389718806,3-2)
(3.26495046379566,3-4)
(3.15475330078091,4-1)
(3.34301135700879,4-2)
(3.48428969428933,4-3)
};
% Format 5
\addplot [black, mark = diamond, mark options = {solid, thick}, only marks, y filter/.code={\pgfmathparse{\pgfmathresult+0.2}\pgfmathresult}] coordinates{
(1.76059952318463,1-2)
(1.86385937553707,1-3)
(1.96969360857475,1-4)
(2.25832742188271,2-1)
(2.51185858411453,2-3)
(2.6472705996547,2-4)
(2.78395943247782,3-1)
(2.94039647233279,3-2)
(3.24997813124171,3-4)
(3.15414539265742,4-1)
(3.27214264487371,4-2)
(3.44205252987905,4-3)
};
\end{axis}

\begin{axis}[
at = {(axis1.south east)},
xshift = 0.125\textwidth,
title = {Second prize},
title style = {font=\small},
width = 0.47\textwidth, 
height = 0.5\textwidth,
xmajorgrids,
ymajorgrids,
%xbar stacked,
%bar width = 10pt,
scaled x ticks = false,
xlabel = {Expected payoff for offensive play},
xlabel style = {align=center, font=\small},
xticklabel style = {/pgf/number format/fixed,/pgf/number format/precision=5},
xmin = 0.4,
xmax = 2.6,
ylabel = {Matches between teams \\ drawn from different pots},
ylabel style = {align=center, font=\small},
symbolic y coords = {1-2,1-3,1-4,2-1,2-3,2-4,3-1,3-2,3-4,4-1,4-2,4-3},
ytick = data,
y dir = reverse,
enlarge y limits = 0.05,
]
% Format 1
\addplot [red, mark = +, mark options = {solid, thick}, only marks] coordinates{
(0.845556195337542,1-2)
(0.674889802058495,1-3)
(0.514285022291528,1-4)
(1.20713311969971,2-1)
(0.808996975458404,2-3)
(0.604599548700197,2-4)
(1.40977230303678,3-1)
(1.20063851945435,3-2)
(0.730320796054002,3-4)
(1.73903873678773,4-1)
(1.48735266822047,4-2)
(1.20078475360184,4-3)
};
% Format 2
\addplot [blue, mark = x, mark options = {solid, thick}, only marks, y filter/.code={\pgfmathparse{\pgfmathresult-0.2}\pgfmathresult}] coordinates{
(0.833413775409095,1-2)
(0.804165592225166,1-3)
(0.796337796294096,1-4)
(1.05826543140977,2-1)
(1.04276383923649,2-3)
(1.03281170377582,2-4)
(1.40040978778115,3-1)
(1.41113264731736,3-2)
(1.41246266609776,3-4)
(2.07984077042085,4-1)
(2.12291291381539,4-2)
(2.12695334476726,4-3)
};
% Format 3
\addplot [brown, mark = otimes, mark options = {solid, thick}, only marks, y filter/.code={\pgfmathparse{\pgfmathresult+0.2}\pgfmathresult}] coordinates{
(0.967339100470249,1-2)
(0.953710428176982,1-3)
(0.957503285425238,1-4)
(1.18009108054651,2-1)
(1.16927206793849,2-3)
(1.18232877794454,2-4)
(1.44351999967102,3-1)
(1.44386328871079,3-2)
(1.45670383875535,3-4)
(1.82207866992512,4-1)
(1.82025555635545,4-2)
(1.8176926388291,4-3)
};
% Format 4
\addplot [ForestGreen, mark = Mercedes star, mark options = {solid, thick}, only marks, y filter/.code={\pgfmathparse{\pgfmathresult-0.2}\pgfmathresult}] coordinates{
(1.09102452776992,1-2)
(1.08899524561674,1-3)
(1.10809819160458,1-4)
(1.35356142931715,2-1)
(1.36496884219473,2-3)
(1.40307268310913,2-4)
(1.6598933389875,3-1)
(1.66289804105923,3-2)
(1.73722710411029,3-4)
(2.08475373900161,4-1)
(2.09729779964251,4-2)
(2.1208028441903,4-3)
};
% Format 5
\addplot [black, mark = diamond, mark options = {solid, thick}, only marks, y filter/.code={\pgfmathparse{\pgfmathresult+0.2}\pgfmathresult}] coordinates{
(1.12671415521255,1-2)
(1.12343823971346,1-3)
(1.1513338394733,1-4)
(1.40624008833861,2-1)
(1.40679575539715,2-3)
(1.45186433241543,2-4)
(1.71195707645682,3-1)
(1.71249473772067,3-2)
(1.79517699393583,3-4)
(2.14011658280104,4-1)
(2.14509172726687,4-2)
(2.17112197579685,4-3)
};
\end{axis}
\end{tikzpicture}
\end{subfigure}

\captionsetup{justification=centerfirst}
\caption{Incentives for offensive play in alternative UCL formats by seeding pots \\ \vspace{0.2cm}
\footnotesize{\emph{Notes:} The tournament formats are detailed in Table~\ref{Table3}. \\
Matches between teams from different seeding pots are distinguished. \\
In each pair $a$-$b$, the first team drawn from the higher-ranked pot $a$ plays at home (top panel) or away (bottom panel). \\
Values are based on Equation~\ref{eq:incentives}.}}
\label{Fig5}
\end{figure}

%\end{document}                                                                                                                                                                

Figure~\ref{Fig5} shows the incentives for offensive play for the first and second prize in the five formats; Table~\ref{Table_D1} in Appendix~\ref{Appendix_D} contains the underlying values.
The difference between Formats~1 and~2 reflects the evolution of incentives when the group stage is replaced by an incomplete round-robin league with the same number of teams and matches, such that the same proportion of teams receives the two prizes. Considering the first prize, all teams have stronger incentives to play offensively in the incomplete round-robin format; the increase ranges from 17\% (a team from Pot~1 plays at home against a team from Pot~2) to 71\% (a team from Pot~4 plays away against a team from Pot~3). For the second prize, the sign of the change in incentives depends on the participating teams. We find stronger incentives in all matches of Pot~4 teams, as well as in almost all matches of Pot~3 teams, except for a slight decrease in a home match against a team from Pot~1. Teams from Pots~1 and 2 also have stronger incentives in all matches, apart from a match played between themselves.

Format~3 differs from Format~2 by adding four teams and two matches per team, against opponents drawn from the same seeding pot. For the first prize, a slight increase (decrease) is found in incentives for offensive play in all matches of Pot~1 (Pot~3) teams, while the implications for Pot~2 teams remain similar. However, the incentives for offensive play regarding the first prize are consistently reduced for Pot~4 teams when these highly balanced matches are played in the tournament. Results for the second prize coincide.

Reducing the proportion of teams receiving the first (second) prize from 50\% (75\%) to 22.2\% (66.7\%) again has a strong impact on incentives to attack. For the first prize, we observe robust increases across all teams and matches, up to 54\% (away matches between teams from Pot~3 and Pot~4). The same pattern with a weaker magnitude (up to 19\%) can be seen for the second prize.

Finally, the difference between Formats~4 and~5 shows that, while the proportion of draws can considerably affect qualification prospects \citep{WinkelmannDeutscherMichels2026}, the estimated parameters of the underlying prediction model have negligible implications on the incentives for offensive play, especially for the first prize.

\begin{table}[t!]
    \centering
    \caption{Average incentives for offensive play in alternative UCL formats}
    \label{Table4}
    \begin{threeparttable}
    \rowcolors{3}{gray!20}{}
    \begin{tabularx}{\textwidth}{l CCCC} \toprule \hiderowcolors
    Team  & \multicolumn{2}{c}{Home} & \multicolumn{2}{c}{Away} \\ 
    Prize & First & Second & First & Second \\ \bottomrule \showrowcolors
    Format 1 & 1.52 & 1.33 & 1.19 & 1.04 \\
    Format 2 & 2.12 ($+$39.4\%) & 1.96 ($+$47.5\%) & 1.46 ($+$22.7\%) & 1.34 ($+$29.8\%) \\
    Format 3 & 1.94 \textcolor{gray!20}{1}($-$8.6\%) & 1.84 \textcolor{gray!20}{1}($-$6.2\%) & 1.42 \textcolor{gray!20}{1}($-$2.2\%) & 1.35 \textcolor{gray!20}{1}($+$0.6\%) \\
    Format 4 & 2.80 ($+$44.4\%) & 2.67 ($+$45.4\%) & 1.65 ($+$16.0\%) & 1.56 ($+$15.8\%) \\
    Format 5 & 2.78 \textcolor{gray!20}{1}($-$0.7\%) & 2.65 \textcolor{gray!20}{1}($-$0.7\%) & 1.70 \textcolor{gray!20}{1}($+$3.0\%) & 1.61 \textcolor{gray!20}{1}($+$3.0\%) \\ \toprule
    \end{tabularx}
\begin{tablenotes} \footnotesize
\item
\textit{Notes:}
The tournament formats are detailed in Table~\ref{Table3}.
\item
Values are averages taken over the 12 pairs of different seeding pots based on Equation~\ref{eq:incentives}. Numbers in parentheses show the changes compared to the format in the above row.
\end{tablenotes}
\end{threeparttable}
\end{table}

Table~\ref{Table4} reports the average incentives for offensive play and their changes as a result of each reform in the UCL design (including the change of the sample used to estimate model parameters), separately for the two prizes and the field of the match. The new format creates stronger incentives to attack, mainly due to the introduction of the incomplete round-robin league and the more exclusive set of prizes. Adding matches between teams drawn from the same seeding pot and increasing the number of participants have marginal effects, similar to the sample used for model calibration.
Since restricting the set of qualifying teams encourages offensive play simultaneously with an increased risk of early elimination, replacing the group stage with an incomplete round-robin league offers a promising way to create stronger incentives without adverse consequences.

\section{Discussion} \label{Sec5}
%\label{sec:remarks}
%\input{files_DW/05_Discussion_DW}

In this paper, we have developed a general simulation framework to quantify incentives for offensive play in a sports tournament by comparing the potential gain from scoring a goal with the potential loss from conceding one using a 0-0 draw (the initial scoreline) as the reference point. The expected payoffs are measured by changes in the probability of reaching the critical thresholds, determined by the qualification rules, in the final ranking. The proposed model is applied to the old group stage and new incomplete round-robin formats of the UEFA Champions League to explore the impact of the major reforms in the 2024/25 season. 

A considerable change is reported in the probabilities of winning the first and second prizes in the new incomplete round-robin format. While these probabilities clearly decrease for the first prize for all teams, regardless of the seeding pot from which they are drawn, the second prize shows a large decline mainly for teams drawn from the weakest Pot~4.

Crucially, the new design is shown to strongly encourage offensive play, as evidenced by the currently available, albeit limited, empirical data. In particular, for the first prize (direct qualification for the Round of 16), our incentive measure increases by 58--155\% across the 12 matches, about 100\% on average. This is especially important for the strongest teams drawn from Pot~1, whose incentives more than double and robustly exceed one in the new format.
Incentives for the second prize (to avoid elimination from European competitions) also rise in all matches, by about 59\% on average.

We have also attempted to decompose the effects of this complex reform package. According to our findings, roughly half of the increase in incentives is due to the more exclusive definition of the two prizes. However, the other half can be directly attributed to the incomplete round-robin format, which appears to be more effective at supporting offensive play than the traditional group stage.

The result is intuitive.
To see why, compare Formats~1 and~2: the former consists of eight groups of four teams playing a double round-robin contest, whereas the latter is an incomplete round-robin tournament involving 32 teams and six rounds. In both formats, a team receives three points for a win and one point for a draw. However, a match also has an \emph{indirect} effect on the points scored by the other teams. Holding all other results fixed, playing a draw rather than losing decreases the average points of the other teams in a group by 2/3 in Format~1. In contrast, the corresponding decrease in Format~2 across the other 31 teams is only 2/31. Thus, although a draw earns the team one additional point relative to a loss in both formats, its indirect benefit is much smaller in Format~2. Consequently, when teams evaluate their points relative to the average points of the other teams, which is necessary to estimate the probability of achieving a prize, the benefit of securing a draw rather than risking a loss diminishes relative to the benefit of winning. This increases the incentive to play offensively in Format~2 compared to Format~1.
%Denote the proportion of draws in the other matches by $\alpha_1$ and $\alpha_2$, respectively.
%In Format 1, if team $i$ loses a particular match, the average number of points scored by the other three teams is between $\left( 3 \times 12 - 5 \times 3 - 11 \alpha_1 \right) / 3 = 7 - 11 \alpha_1 / 3$ and $\left( 3 \times 12 - 11 \alpha_1 \right) / 3 = 12 - 11 \alpha_1 / 3$, depending on the results of other matches played by team $i$. If team $i$ plays a draw (wins) in this particular match, both the lower and upper bounds decrease by 2/3 (1).
%In Format 2, if team $i$ loses a particular match, the average number of points scored by the other 31 teams is between $\left( 3 \times 96 - 5 \times 3 - 95 \alpha_2 \right) / 31 = 273/31 - 95 \alpha_2 / 31$ and $\left( 3 \times 96 - 95 \alpha_2 \right) / 31 = 273/31 - 95 \alpha_2 / 31$, depending on the results of other matches played by team $i$. If team $i$ plays a draw (wins) in this particular match, both the lower and upper bounds decrease by 2/31 (3/31).

Our main message is important for tournament organisers, especially in football, where draws are quite frequent and can be achieved by unattractive, destructive play. Indeed, the rule of three points for a win, probably the most important change in the recent history of football, was introduced with the explicit aim of encouraging more attacking play. Since this implication is not so clear in practice as we have seen in Section~\ref{Sec2}, replacing the group stage with a single incomplete round-robin league is worth considering as an alternative or complement.

Naturally, a careful interpretation of the numerical results is warranted. The simulations use static Elo ratings as a proxy for team strengths and ignore the order of matches, despite some evidence in the sports economics literature on the impact of the match schedule \citep{KrumerLechner2017}, including in the UCL groups \citep{Krumer2021a}. The kick-off time of a match is also neglected, while it might affect home advantage \citep{Krumer2020a}.
Furthermore, taking the average value of $\mathcal{I}_{jk}$ is realistic only if the teams do not know the results of any other matches. However, in later rounds, the outcomes of earlier matches become common knowledge, and the average value of incentives is no longer valid. This holds particularly for matches played in the last round(s), which may easily become non-competitive \citep{ChaterArrondelGayantLaslier2021, CsatoMolontayPinter2024, DevriesereGoossensSpieksma2026}.
Finally, Equation~\eqref{eq:incentives} assumes equal risks of scoring and conceding a goal for each team. Multiplying the formula by a homogeneous constant is unlikely to alter the preference between the two formats, but the weights of these risks may be heterogeneous: a stronger team could have a higher probability of scoring a goal and/or a lower probability of conceding a goal when playing offensively. Estimating these parameters requires further research and is clearly beyond the scope of the current study.

To conclude, our paper suggests several promising directions for future studies. Sensitivity analyses can be conducted with alternative model formulations and prize definitions to explore the robustness of the results. Indeed, the group winners enjoyed some advantages over the runners-up in the group stage of the UEFA Champions League, and a better rank in the league phase implies, on average, weaker opponents and occasionally the opportunity to play the second leg at home in the subsequent knockout stages. The proposed model can also be applied to measure incentives to attack in other tournaments and sports. From a managerial perspective, the proposed approach could help determine the optimal qualification thresholds with respect to offensive play for a given (sub)set of teams, which would be crucial for promoting sports and generating more revenue. Finally, the quantification of incentives adds a novel dimension to attractiveness, a well-established criterion for comparing and evaluating tournament designs. 

\subsection*{Acknowledgements}
\addcontentsline{toc}{section}{Acknowledgements}
\noindent
We are grateful to \emph{Aleksei Y.~Kondratev} and \emph{Christian Deutscher} for useful remarks.

\subsection*{Data availability statement}
\addcontentsline{toc}{section}{Data availability statement}
\noindent
The data used in the study are publicly available. Football Club Elo Ratings are retrieved from \url{www.clubelo.com}. The official match schedules and results for the UEFA Champions League seasons are obtained from UEFA.

\subsection*{Generative AI statement} 
\addcontentsline{toc}{section}{Generative AI statement}
\noindent
During the preparation of this work, the authors used ChatGPT-5 to assist with improving language and clarity. The R code was written by the second author from scratch, and later refined with the help of ChatGPT-5 to enhance efficiency and readability. The authors reviewed and edited all content to ensure it meets the required standards and take full responsibility for the final version of this publication.

\subsection*{Funding information}
\addcontentsline{toc}{section}{Funding information}
\noindent
The research was supported by the National Research, Development and Innovation Office under Grants Advanced 152220 and FK 145838, and the J\'anos Bolyai Research Scholarship of the Hungarian Academy of Sciences.

\bibliographystyle{apalike}
\bibliography{All_references}

%\newpage
%\label{sec:appendix}
%\input{files_DW/06_Appendix}

\clearpage
\section*{Appendix}
\addcontentsline{toc}{section}{Appendix}

\appendix
\begin{appendices}

\counterwithin{table}{section}
\counterwithin{figure}{section}

\section{Probabilities of achieving the two prizes} \label{Appendix_A}

\begin{table}[ht!]
    \centering
    \caption{Probability of obtaining the first prize in the old and new UCL formats}
    \label{Table_A1}
\begin{threeparttable} 
\rowcolors{3}{gray!20}{}
    \begin{tabularx}{\textwidth}{Ccc CCC CCC} \toprule \hiderowcolors
        \multirow{2}{*}{Team} & \multirow{2}{*}{Opponent} & \multirow{2}{*}{Location} & \multicolumn{3}{c}{Old format} & \multicolumn{3}{c}{New format} \\
          & & & Win & Draw & Loss & Win & Draw & Loss \\ \bottomrule \showrowcolors
        1 & 2 & 1 & 92.18 & 82.16 & 72.20 & 60.77 & 42.18 & 32.23 \\
        1 & 2 & 0 & 94.29 & 86.14 & 76.78 & 66.42 & 48.19 & 37.84 \\
        1 & 3 & 1 & 91.46 & 80.34 & 67.43 & 57.69 & 39.56 & 30.24 \\
        1 & 3 & 0 & 93.77 & 85.06 & 73.36 & 62.99 & 44.67 & 34.85 \\
        1 & 4 & 1 & 89.46 & 78.97 & 67.61 & 55.15 & 37.48 & 28.74 \\
        1 & 4 & 0 & 91.47 & 82.83 & 72.78 & 59.58 & 41.43 & 32.21 \\ \hline
        2 & 1 & 1 & 77.06 & 60.83 & 50.49 & 35.50 & 19.55 & 12.97 \\
        2 & 1 & 0 & 81.08 & 65.76 & 54.75 & 41.15 & 24.26 & 16.78 \\
        2 & 3 & 1 & 75.17 & 54.44 & 35.64 & 28.26 & 14.88 &  \textcolor{gray!20}{0}9.79 \\
        2 & 3 & 0 & 80.72 & 62.84 & 43.45 & 33.33 & 18.33 & 12.36 \\
        2 & 4 & 1 & 70.15 & 52.37 & 37.26 & 25.91 & 13.50 &  \textcolor{gray!20}{0}8.93 \\
        2 & 4 & 0 & 74.49 & 58.60 & 43.75 & 29.96 & 16.05 & 10.79 \\ \hline
        3 & 1 & 1 & 54.49 & 35.47 & 25.58 & 17.73 &  \textcolor{gray!20}{0}7.92 &  \textcolor{gray!20}{0}4.64 \\
        3 & 1 & 0 & 60.40 & 40.94 & 29.30 & 21.76 & 10.53 &  \textcolor{white}{0}6.50 \\
        3 & 2 & 1 & 54.57 & 32.10 & 17.86 & 14.80 &  \textcolor{gray!20}{0}6.40 &  \textcolor{gray!20}{0}3.71 \\
        3 & 2 & 0 & 62.30 & 40.24 & 23.24 & 18.66 &  \textcolor{white}{0}8.61 &  \textcolor{white}{0}5.19 \\
        3 & 4 & 1 & 44.37 & 28.09 & 17.95 & 11.47 &  \textcolor{gray!20}{0}4.84 &  \textcolor{gray!20}{0}2.85 \\
        3 & 4 & 0 & 49.12 & 33.02 & 22.17 & 14.01 &  \textcolor{white}{0}6.09 &  \textcolor{white}{0}3.65 \\ \hline
        4 & 1 & 1 & 31.40 & 17.15 & 11.27 &  8.24 &  \textcolor{gray!20}{0}3.50 &  \textcolor{gray!20}{0}2.08 \\
        4 & 1 & 0 & 36.35 & 20.71 & 13.35 & 10.35 &  \textcolor{white}{0}4.70 &  \textcolor{white}{0}2.91 \\
        4 & 2 & 1 & 30.74 & 15.17 &  \textcolor{gray!20}{0}7.68 &  \textcolor{gray!20}{0}6.74 &  \textcolor{gray!20}{0}2.82 &  \textcolor{gray!20}{0}1.68 \\
        4 & 2 & 0 & 36.94 & 20.06 & 10.42 &  \textcolor{white}{0}8.69 &  \textcolor{white}{0}3.80 &  \textcolor{white}{0}2.30 \\
        4 & 3 & 1 & 26.75 & 13.80 &  \textcolor{gray!20}{0}7.32 &  \textcolor{gray!20}{0}5.80 &  \textcolor{gray!20}{0}2.42 &  \textcolor{gray!20}{0}1.46 \\
        4 & 3 & 0 & 31.59 & 17.71 &  \textcolor{white}{0}9.86 &  \textcolor{white}{0}7.47 &  \textcolor{white}{0}3.19 &  \textcolor{white}{0}1.95 \\ \bottomrule      
    \end{tabularx}
\begin{tablenotes} \footnotesize
\item
\textit{Notes:}
Columns \textit{Team} and \textit{Opponent} indicate the respective seeding pots. Column \textit{Location} is a binary variable taking the value 1 if the team plays at home and 0 if it plays away.
\item
First prize is qualification for the Round of 16 in the old and direct qualification for the Round of 16 in the new format.
\item
Probabilities are given in \%.
\end{tablenotes}
\end{threeparttable}
\end{table}

\begin{table}[t!]
    \centering
    \caption{Probability of obtaining the second prize in the old and new UCL formats}
    \label{Table_A2}
\begin{threeparttable} 
\rowcolors{3}{gray!20}{}
    \begin{tabularx}{\textwidth}{Ccc CCC CCC} \toprule \hiderowcolors
        \multirow{2}{*}{Team} & \multirow{2}{*}{Opponent} & \multirow{2}{*}{Location} & \multicolumn{3}{c}{Old format} & \multicolumn{3}{c}{New format} \\
          & & & Win & Draw & Loss & Win & Draw & Loss \\ \bottomrule \showrowcolors
        1 & 2 & 1 & 98.42 & 95.70 & 92.83 & 97.01 & 92.15 & 88.04 \\
        1 & 2 & 0 & 98.89 & 96.74 & 94.21 & 97.84 & 94.05 & 90.68 \\
        1 & 3 & 1 & 98.39 & 95.10 & 90.93 & 96.38 & 90.78 & 86.11 \\
        1 & 3 & 0 & 98.96 & 96.54 & 92.95 & 97.45 & 93.08 & 89.19 \\
        1 & 4 & 1 & 98.35 & 94.35 & 87.82 & 95.56 & 89.34 & 84.26 \\
        1 & 4 & 0 & 99.03 & 96.29 & 90.96 & 96.89 & 91.80 & 87.37 \\ \hline
        2 & 1 & 1 & 94.47 & 88.42 & 83.68 & 90.74 & 79.70 & 72.15 \\
        2 & 1 & 0 & 95.54 & 90.13 & 85.65 & 92.43 & 82.85 & 76.04 \\
        2 & 3 & 1 & 94.32 & 85.53 & 76.39 & 87.12 & 73.50 & 64.45 \\
        2 & 3 & 0 & 96.16 & 89.17 & 80.52 & 90.21 & 78.47 & 70.13 \\
        2 & 4 & 1 & 94.33 & 83.61 & 69.34 & 84.94 & 70.53 & 61.20 \\
        2 & 4 & 0 & 96.47 & 88.70 & 75.80 & 88.47 & 75.59 & 66.71 \\ \hline
        3 & 1 & 1 & 84.88 & 74.07 & 66.82 & 76.88 & 60.09 & 50.68 \\
        3 & 1 & 0 & 87.09 & 76.81 & 69.51 & 79.82 & 64.32 & 55.26 \\
        3 & 2 & 1 & 84.90 & 71.89 & 62.39 & 73.95 & 55.99 & 46.09 \\
        3 & 2 & 0 & 88.18 & 76.22 & 66.27 & 78.10 & 61.51 & 51.82 \\
        3 & 4 & 1 & 85.54 & 66.69 & 46.23 & 67.49 & 48.76 & 38.88 \\
        3 & 4 & 0 & 90.29 & 75.35 & 54.86 & 73.06 & 54.81 & 44.65 \\ \hline
        4 & 1 & 1 & 60.94 & 45.00 & 36.41 & 54.92 & 37.11 & 29.09 \\
        4 & 1 & 0 & 64.89 & 48.59 & 39.24 & 58.42 & 40.91 & 32.72 \\
        4 & 2 & 1 & 61.95 & 42.42 & 31.26 & 51.63 & 33.55 & 25.59 \\
        4 & 2 & 0 & 68.27 & 48.58 & 35.38 & 56.45 & 38.41 & 30.00 \\
        4 & 3 & 1 & 63.21 & 39.70 & 24.04 & 48.34 & 30.54 & 22.83 \\
        4 & 3 & 0 & 71.33 & 48.72 & 29.89 & 54.09 & 35.86 & 27.46 \\ \bottomrule        
    \end{tabularx}
\begin{tablenotes} \footnotesize
\item
\textit{Notes:}
Columns \textit{Team} and \textit{Opponent} indicate the respective seeding pots. Column \textit{Location} is a binary variable taking the value 1 if the team plays at home and 0 if it plays away.
\item
Second prize is avoiding elimination from UEFA club competitions in both formats.
\item
Probabilities are given in \%.
\end{tablenotes}
\end{threeparttable}
\end{table}

\clearpage
\counterwithin{table}{section}
\counterwithin{figure}{section}

\section{Robustness to the reference scoreline} \label{Appendix_B}
%\label{sec:appx_further_scoreline}

Each football match starts with the scoreline of 0-0. Therefore, we use this benchmark to assess the incentives for offensive play in the baseline analysis. However, it is also useful to measure the incentives from scorelines of 1-1 and 2-2, after both teams have scored an equal number of goals.

\begin{table}[ht!]
    \centering
    \caption{Incentives for offensive play in the new \\ UCL format under different scorelines by seeding pots}
    \label{Table_B1}
\begin{threeparttable}
\rowcolors{3}{gray!20}{}
    \begin{tabularx}{\textwidth}{Ccc CCC CCC} \toprule \hiderowcolors
        \multirow{2}{*}{Team} & \multirow{2}{*}{Opponent} & \multirow{2}{*}{Location} & \multicolumn{3}{c}{First prize} & \multicolumn{3}{c}{Second prize} \\
        & & & 0-0 & 1-1 & 2-2 & 0-0 & 1-1 & 2-2 \\ \bottomrule \showrowcolors
        1 & 2 & 1 & 1.87 & 1.87 & 1.87 & 1.18 & 1.18 & 1.18 \\
        1 & 2 & 0 & 1.76 & 1.76 & 1.76 & 1.13 & 1.13 & 1.13 \\
        1 & 3 & 1 & 1.95 & 1.95 & 1.95 & 1.20 & 1.20 & 1.20 \\
        1 & 3 & 0 & 1.86 & 1.87 & 1.87 & 1.12 & 1.12 & 1.12 \\
        1 & 4 & 1 & 2.02 & 2.02 & 2.02 & 1.22 & 1.22 & 1.22 \\
        1 & 4 & 0 & 1.97 & 1.97 & 1.97 & 1.15 & 1.15 & 1.15 \\ \hline
        2 & 1 & 1 & 2.42 & 2.42 & 2.42 & 1.46 & 1.46 & 1.46 \\
        2 & 1 & 0 & 2.26 & 2.26 & 2.26 & 1.41 & 1.41 & 1.40 \\
        2 & 3 & 1 & 2.63 & 2.63 & 2.63 & 1.51 & 1.50 & 1.50 \\
        2 & 3 & 0 & 2.51 & 2.51 & 2.51 & 1.41 & 1.41 & 1.41 \\
        2 & 4 & 1 & 2.72 & 2.72 & 2.72 & 1.54 & 1.54 & 1.54 \\
        2 & 4 & 0 & 2.65 & 2.65 & 2.65 & 1.45 & 1.45 & 1.45 \\ \hline
        3 & 1 & 1 & 2.98 & 2.98 & 2.98 & 1.79 & 1.78 & 1.78 \\
        3 & 1 & 0 & 2.78 & 2.79 & 2.78 & 1.71 & 1.71 & 1.71 \\
        3 & 2 & 1 & 3.12 & 3.12 & 3.12 & 1.81 & 1.81 & 1.81 \\
        3 & 2 & 0 & 2.94 & 2.94 & 2.94 & 1.71 & 1.71 & 1.71 \\
        3 & 4 & 1 & 3.34 & 3.33 & 3.33 & 1.90 & 1.89 & 1.89 \\
        3 & 4 & 0 & 3.25 & 3.25 & 3.25 & 1.80 & 1.79 & 1.79 \\ \hline
        4 & 1 & 1 & 3.32 & 3.31 & 3.31 & 2.22 & 2.22 & 2.22 \\
        4 & 1 & 0 & 3.15 & 3.15 & 3.15 & 2.14 & 2.14 & 2.14 \\
        4 & 2 & 1 & 3.41 & 3.41 & 3.41 & 2.27 & 2.27 & 2.27 \\
        4 & 2 & 0 & 3.27 & 3.27 & 3.27 & 2.15 & 2.14 & 2.14 \\
        4 & 3 & 1 & 3.53 & 3.53 & 3.53 & 2.31 & 2.31 & 2.30 \\
        4 & 3 & 0 & 3.44 & 3.44 & 3.44 & 2.17 & 2.17 & 2.17 \\ \bottomrule   
    \end{tabularx}    
\begin{tablenotes} \footnotesize
\item
\textit{Notes:}
Columns \textit{Team} and \textit{Opponent} indicate the respective seeding pots. Column \textit{Location} is a binary variable taking the value 1 if the team plays at home and 0 if it plays away.
\item
Incentives for offensive play are computed based on Equation~\eqref{eq:incentives}.
\end{tablenotes}
\end{threeparttable}
\end{table}

Table~\ref{Table_B1} reports the incentives for offensive play in all matches between clubs from different seeding pots in the new incomplete round-robin format, both for the first prize (top eight) and second prize (top 24).
For both prizes and all three scorelines considered, the variance in the values can be fully attributed to statistical noise in the underlying simulation model. Hence, incentives for offensive play in the new UCL format of the UCL do not depend on the current scoreline, as long as the result remains a draw.

Naturally, if one team leads, the other has a strong incentive to attack, with the possible exception of some matches at the end of the tournament (cf.\ \citet{CsatoGyimesi2026b}).

\clearpage
\counterwithin{table}{section}
\counterwithin{figure}{section}
\makeatletter
\renewcommand{\thesubfigure}{(\alph{subfigure})}
\renewcommand{\p@subfigure}{\thefigure}
\makeatother

\section{Classification according to Elo difference} \label{Appendix_C}

In Section~\ref{Sec44}, the matches are clustered based on seeding pots of the teams. Now we consider an alternative aggregation according to their exact strength.
For this purpose, the differences between the Elo ratings of the two opposing teams are divided into bins of 100 points and averaged across all matches between teams in these bins. Matches where a draw and a loss are equivalent, as both give the same probability to achieve the prize (i.e.\ the denominator in Equation~\eqref{eq:incentives} would be zero), are excluded from the analysis. 

\begin{figure}[ht!]
    \centering
    \begin{subfigure}[t]{0.49\linewidth}
        \centering
        \caption{Group stage, home team}
        \label{Fig_C1a}
        \includegraphics[width=\linewidth]{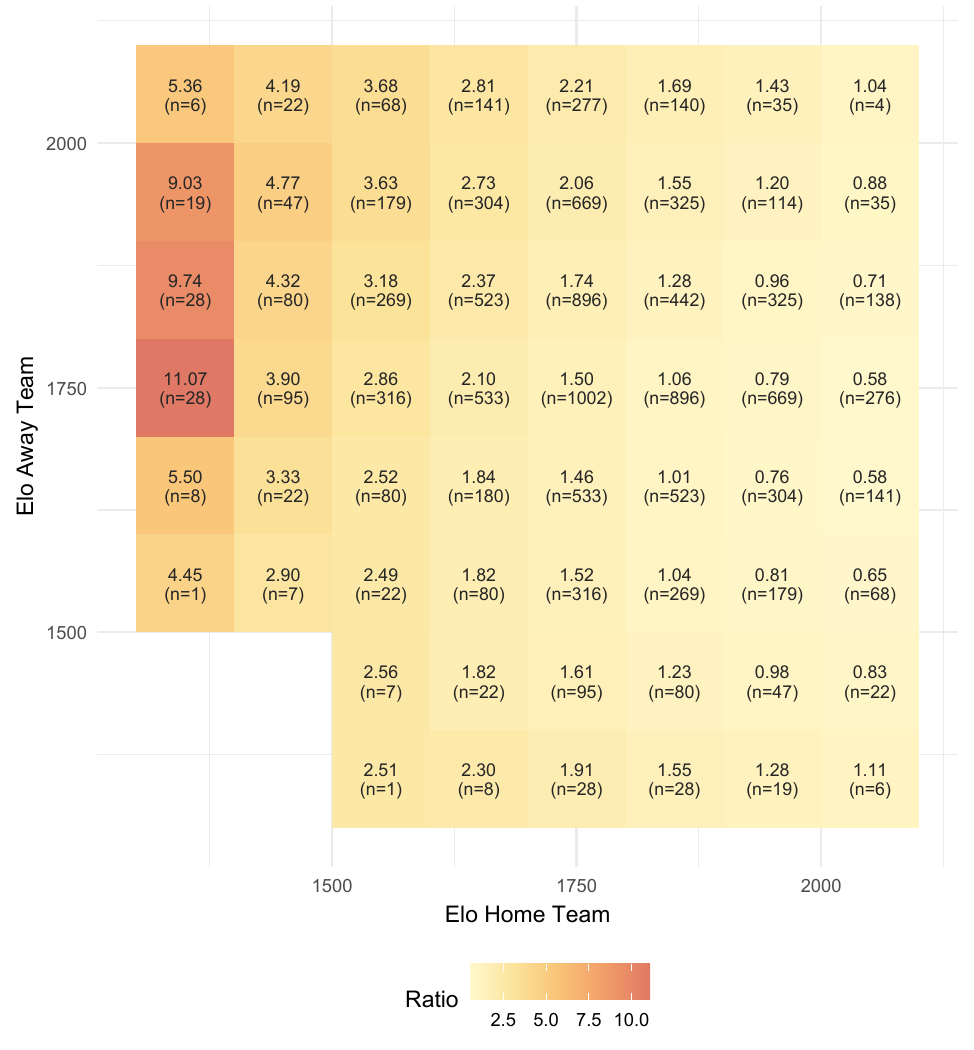}
    \end{subfigure}\hfill
    \begin{subfigure}[t]{0.49\linewidth}
        \centering
        \caption{Incomplete round-robin, home team}
        \label{Fig_C1b}
        \includegraphics[width=\linewidth]{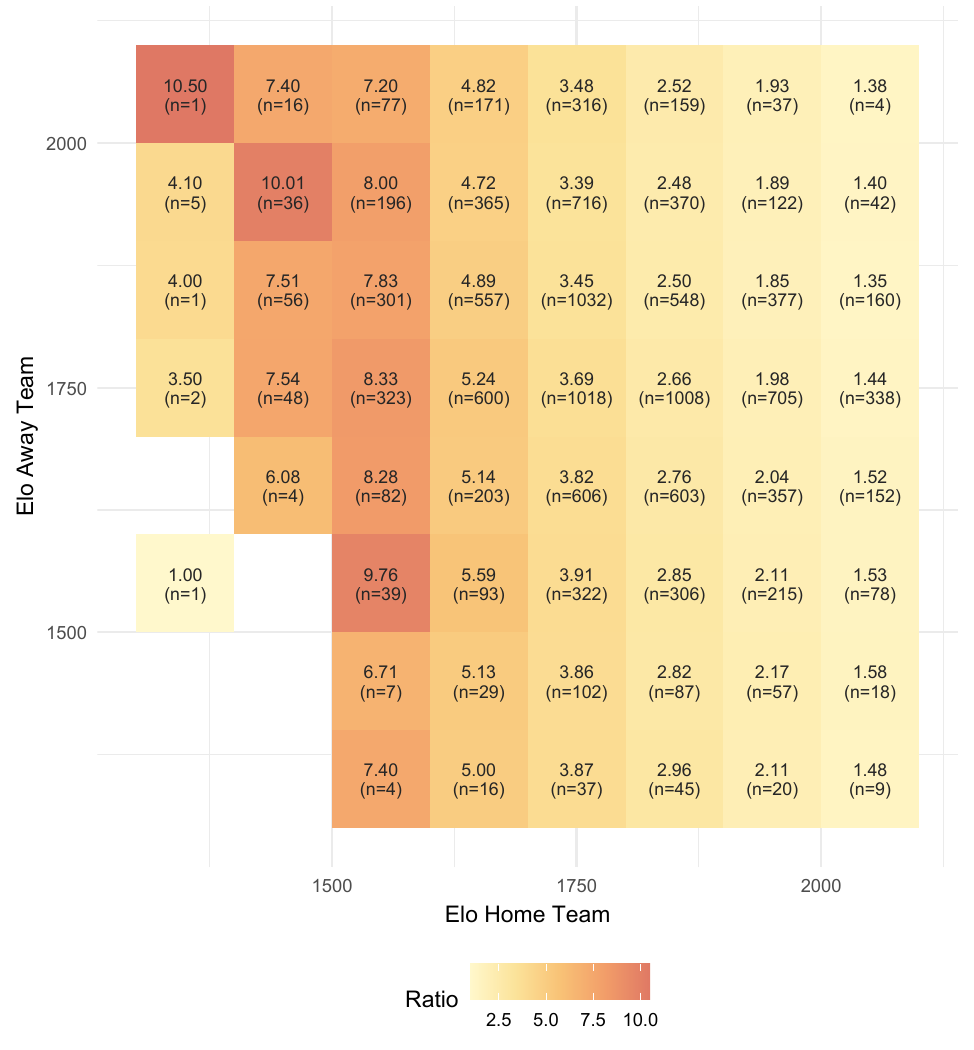}
    \end{subfigure}

\vspace{0.25cm}
    \begin{subfigure}[t]{0.49\linewidth}
        \centering
        \caption{Group stage, away team}
        \label{Fig_C1c}
        \includegraphics[width=\linewidth]{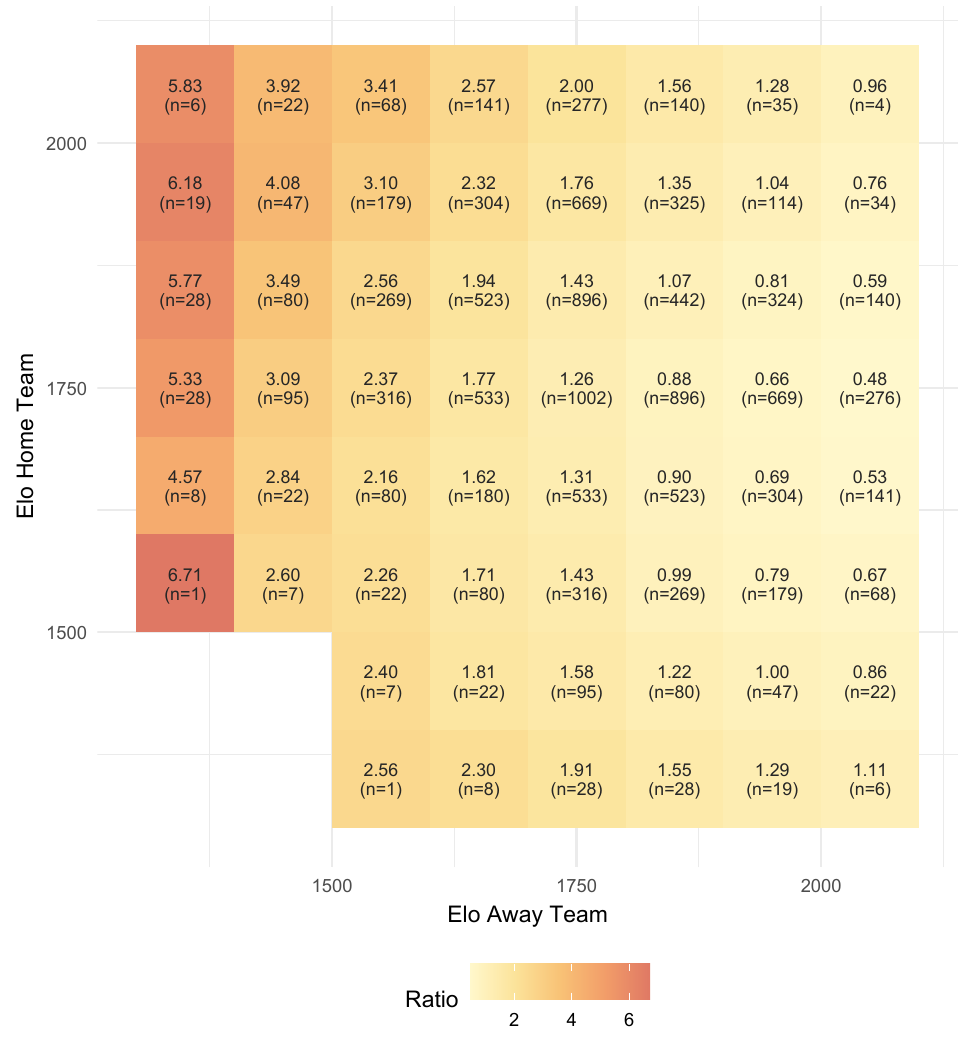}
    \end{subfigure}\hfill
    \begin{subfigure}[t]{0.49\linewidth}
        \centering
        \caption{Incomplete round-robin, away team}
        \label{Fig_C1d}
        \includegraphics[width=\linewidth]{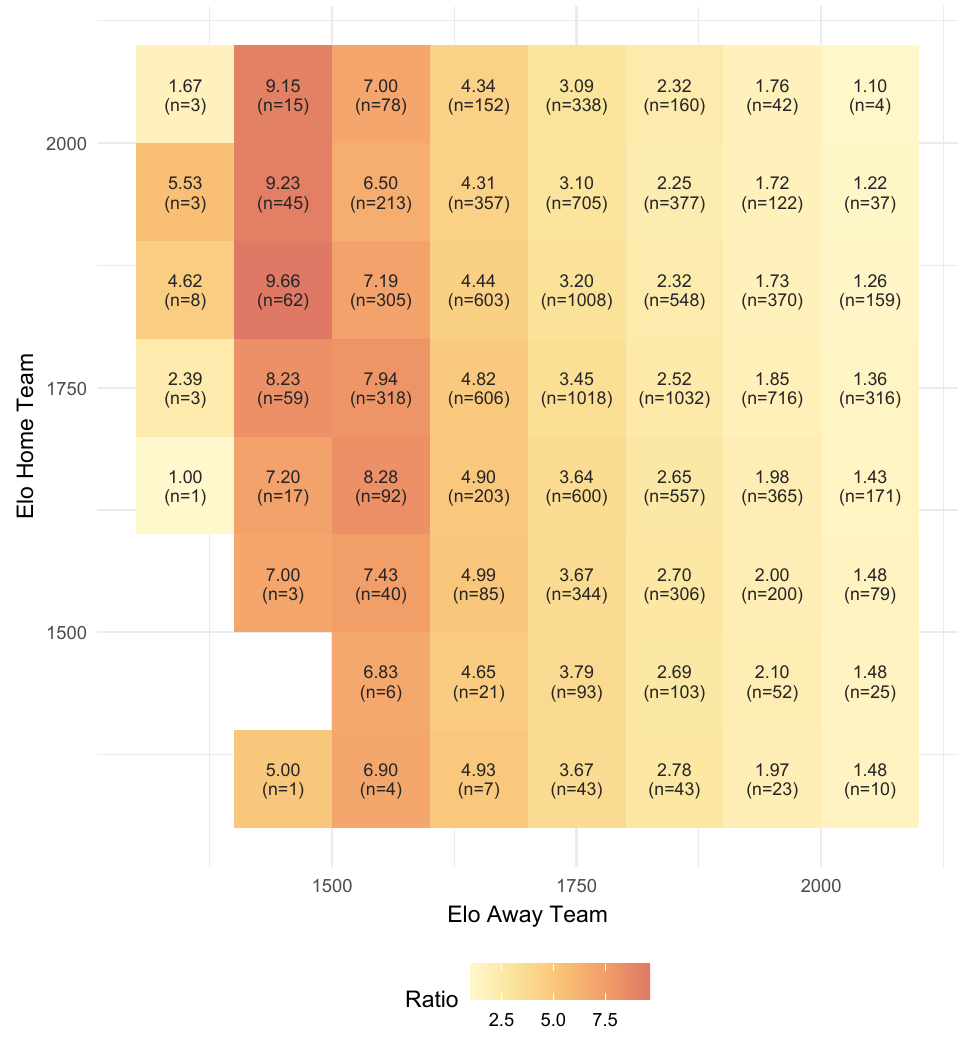}
    \end{subfigure}
\captionsetup{justification=centerfirst}
\caption{Incentives for offensive play in the old and new UCL formats, first prize \\ \vspace{0.2cm}
\footnotesize{
Top Figures~\ref{Fig_C1a} and \ref{Fig_C1b}: Home team is on the x-axis, its opponent (away team) is on the y-axis. \\
Bottom Figures~\ref{Fig_C1c} and \ref{Fig_C1d}: Away team is on the x-axis, its opponent (home team) is on the y-axis. \\
Incentives for offensive play are computed based on Equation~\eqref{eq:incentives}.}
}
\label{Fig_C1}
\end{figure}

The top two panels of Figure~\ref{Fig_C1} compare the incentives of the home team for the first prize in the old (Figure~\ref{Fig_C1a}) and new (Figure~\ref{Fig_C1b}) formats. The strongest incentives arise for teams with the lowest Elo ratings, driven by the low probability of receiving the prize. On the contrary, incentives for teams with higher Elo ratings decrease to 0.58. Regarding the strength of the opponent, we find a $\cup$-shaped effect. Incentives are the lowest when playing against teams with Elo ratings between 1500 and 1800, and the highest when playing against strong or weak teams. This aligns with Figure~\ref{Fig4}, where the incentives for the first prize are the strongest for the home team in the old format when it plays against Pot~1 teams, with the maximum value for teams drawn from Pot~4. 

In the new format, the results are somewhat different. First, the incentives are consistently higher; no value below one occurs. Furthermore, there is less variation in incentives with respect to opponent strength. Still, the incentives increase with lower Elo ratings of the team under consideration, except for a few observations for teams with especially low Elo ratings, which can be attributed partially to the limited sample size in these bins. Again, the results align with those in Figure~\ref{Fig4}, where the seeding pot of the opponent has only a minor impact on the incentives for the first prize in the new format.
On average, the incentives are slightly higher in home matches compared with away matches between teams with the same Elo ratings (Figures~\ref{Fig_C1c} and \ref{Fig_C1d}).

\begin{figure}[t!]
    \centering
    \begin{subfigure}[t]{0.49\linewidth}
        \centering
        \caption{Group stage, home team}
        \label{Fig_C2a}
        \includegraphics[width=\linewidth]{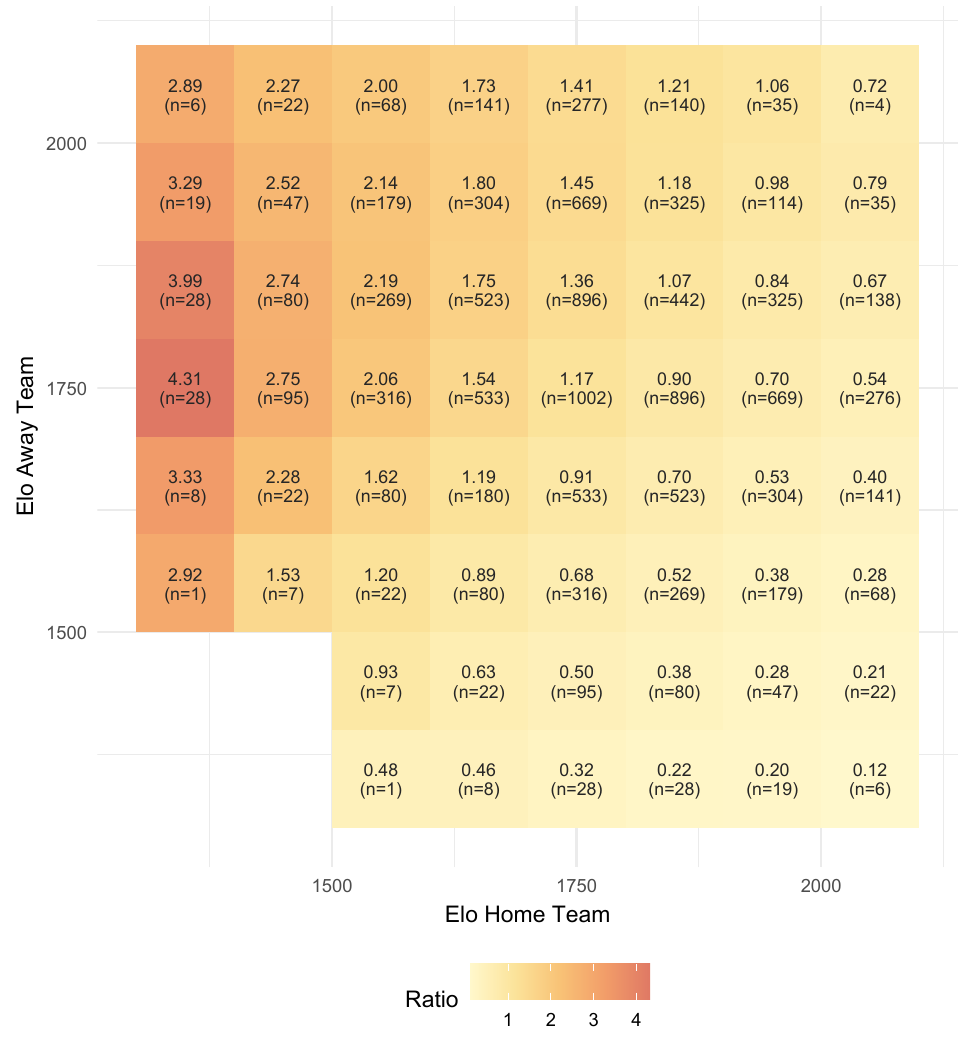}
    \end{subfigure}\hfill
    \begin{subfigure}[t]{0.49\linewidth}
        \centering
        \caption{Incomplete round-robin, home team}
        \label{Fig_C2b}
        \includegraphics[width=\linewidth]{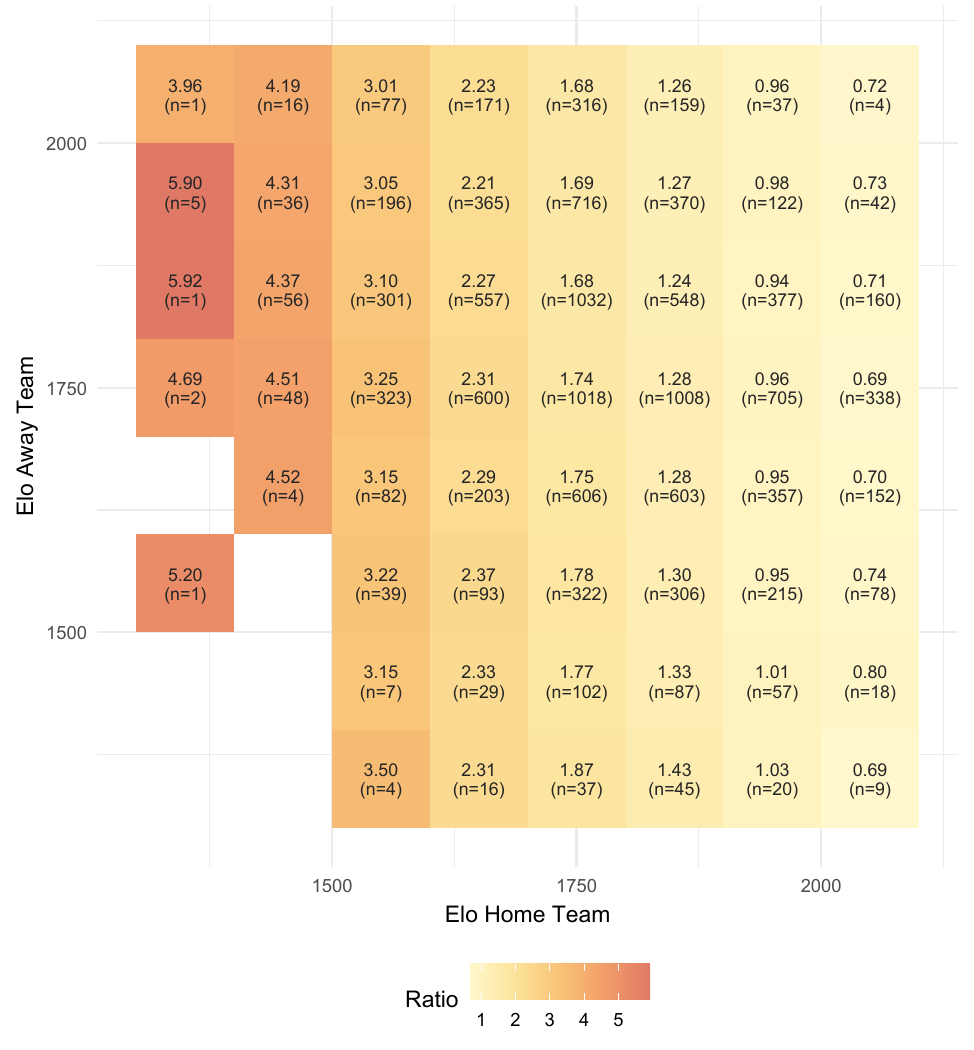}
    \end{subfigure}

\vspace{0.25cm}
    \begin{subfigure}[t]{0.49\linewidth}
        \centering
        \caption{Group stage, away team}
        \label{Fig_C2c}
        \includegraphics[width=\linewidth]{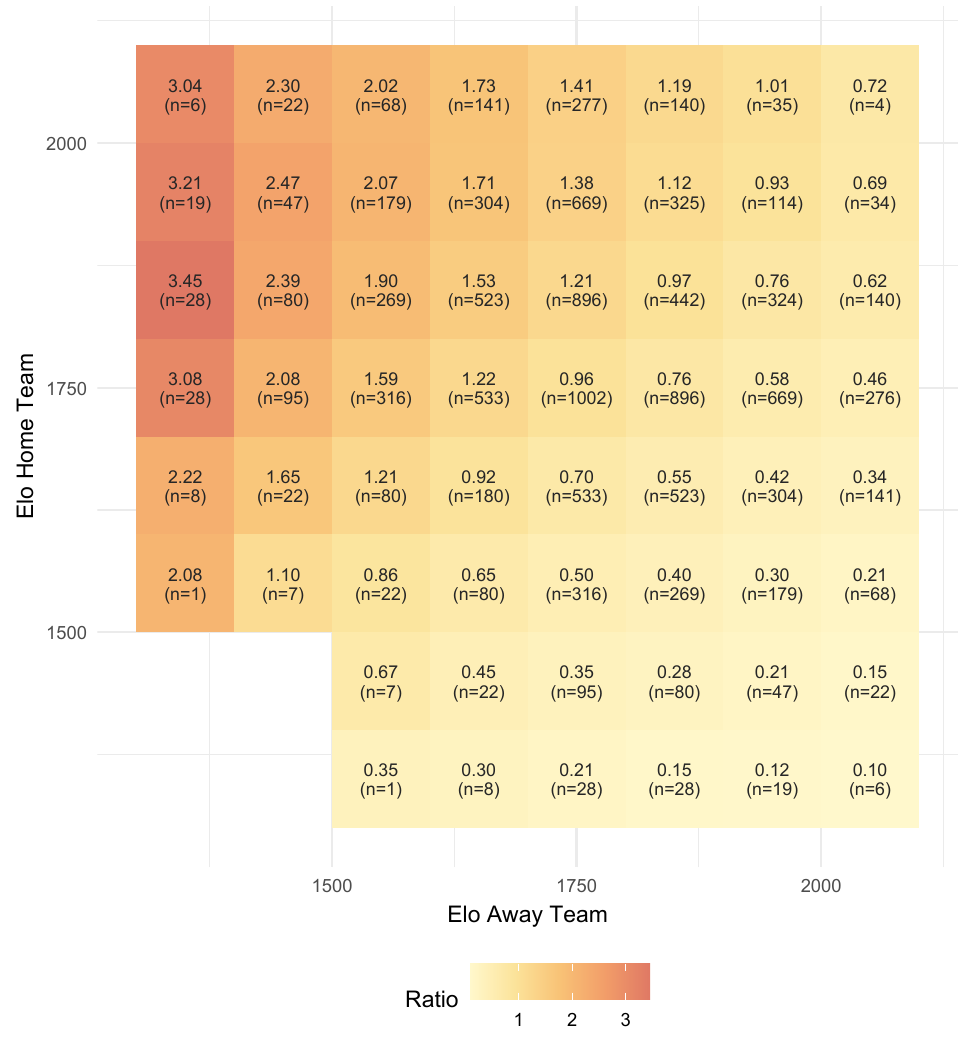}
    \end{subfigure}\hfill
    \begin{subfigure}[t]{0.49\linewidth}
        \centering
        \caption{Incomplete round-robin, away team}
        \label{Fig_C2d}
        \includegraphics[width=\linewidth]{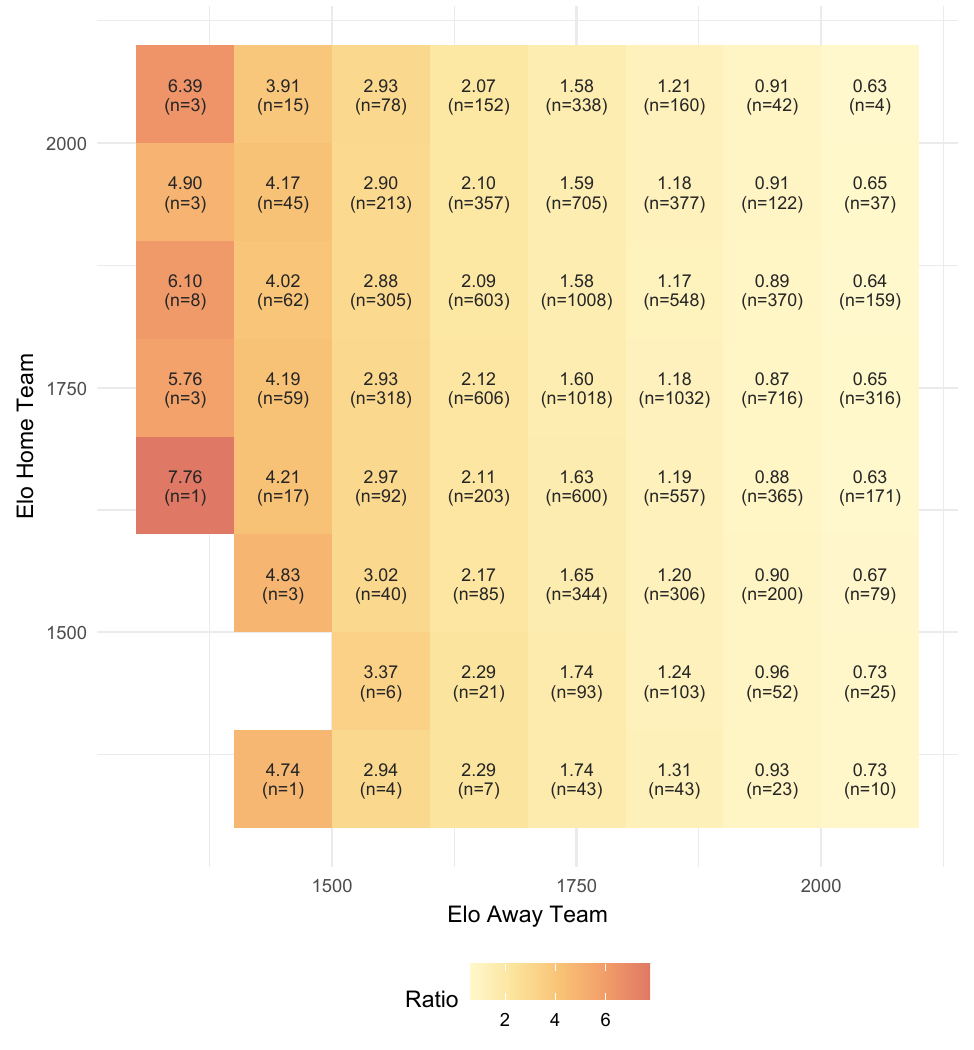}
    \end{subfigure}
\captionsetup{justification=centerfirst}
\caption{Incentives for offensive play in the old and new UCL formats, second prize \\ \vspace{0.2cm}
\footnotesize{
Top Figures~\ref{Fig_C2a} and \ref{Fig_C2b}: Home team is on the x-axis, its opponent (away team) is on the y-axis. \\
Bottom Figures~\ref{Fig_C2c} and \ref{Fig_C2d}: Away team is on the x-axis, its opponent (home team) is on the y-axis. \\
Incentives for offensive play are computed based on Equation~\eqref{eq:incentives}.}}
\label{Fig_C2}
\end{figure}

Figure~\ref{Fig_C2} shows the results for the second prize. In the old format (Figures~\ref{Fig_C2a} and \ref{Fig_C2c}), we find incentives consistently below one against weak opponents, while they increase with the strength of the opposing team. However, for teams with an Elo rating above 2000, the incentives are below one against all opponents. Similar to the first prize, the incentives are stronger for weaker teams under consideration. 
In the new format (Figures~\ref{Fig_C2b} and \ref{Fig_C2d}), there is again only a minimal variation as a function of opponent strength, and the same increase can be observed in incentives for lower Elo ratings of a given team. The incentives remain, on average, higher in home matches than in away matches.

\clearpage
\counterwithin{table}{section}
\counterwithin{figure}{section}

\section{Comparison of incentives in alternative formats} \label{Appendix_D}
%\label{sec:appx_variants}

\begin{table}[ht!]
    \centering
    \caption{Incentives for offensive play in alternative UCL formats by seeding pots}
    \label{Table_D1}
\centerline{
\begin{threeparttable} 
\rowcolors{1}{}{gray!20}
    \begin{tabularx}{1.1\textwidth}{ccc CCCCC}
        Team & Opponent & Location & Format 1 & Format 2 & Format 3 & Format 4 & Format 5 \\ \bottomrule
        1 & 2 & 1 & 1.01 (0.95) & 1.18 (0.88) & 1.32 (1.01) & 1.87 (1.14) & 1.87 (1.18) \\
        1 & 2 & 0 & 0.87 (0.85) & 1.20 (0.87) & 1.25 (0.97) & 1.77 (1.09) & 1.76 (1.13) \\
        1 & 3 & 1 & 0.86 (0.79) & 1.27 (0.86) & 1.34 (1.01) & 1.95 (1.16) & 1.95 (1.20) \\
        1 & 3 & 0 & 0.74 (0.67) & 1.10 (0.83) & 1.27 (0.95) & 1.87 (1.09) & 1.86 (1.12) \\
        1 & 4 & 1 & 0.92 (0.61) & 1.11 (0.80) & 1.40 (1.03) & 2.02 (1.18) & 2.02 (1.22) \\
        1 & 4 & 0 & 0.86 (0.51) & 1.20 (0.80) & 1.33 (0.96) & 1.96 (1.11) & 1.97 (1.15) \\ \hline
        2 & 1 & 1 & 1.57 (1.28) & 1.66 (1.12) & 1.68 (1.22) & 2.43 (1.41) & 2.42 (1.46) \\
        2 & 1 & 0 & 1.39 (1.21) & 1.74 (1.14) & 1.60 (1.18) & 2.28 (1.35) & 2.26 (1.41) \\
        2 & 3 & 1 & 1.10 (0.96) & 1.83 (1.16) & 1.74 (1.25) & 2.63 (1.46) & 2.63 (1.51) \\
        2 & 3 & 0 & 0.92 (0.81) & 1.54 (1.06) & 1.65 (1.17) & 2.53 (1.36) & 2.51 (1.41) \\
        2 & 4 & 1 & 1.18 (0.75) & 1.58 (1.04) & 1.81 (1.28) & 2.72 (1.49) & 2.72 (1.54) \\
        2 & 4 & 0 & 1.07 (0.60) & 1.73 (1.03) & 1.72 (1.18) & 2.64 (1.40) & 2.65 (1.45) \\ \hline
        3 & 1 & 1 & 1.92 (1.49) & 2.25 (1.48) & 2.04 (1.49) & 3.01 (1.73) & 2.98 (1.79) \\
        3 & 1 & 0 & 1.67 (1.41) & 2.34 (1.53) & 1.94 (1.44) & 2.80 (1.66) & 2.78 (1.71) \\
        3 & 2 & 1 & 1.58 (1.37) & 2.51 (1.55) & 2.08 (1.51) & 3.14 (1.76) & 3.12 (1.81) \\
        3 & 2 & 0 & 1.30 (1.20) & 2.05 (1.40) & 1.96 (1.44) & 2.97 (1.66) & 2.94 (1.71) \\
        3 & 4 & 1 & 1.60 (0.92) & 2.10 (1.41) & 2.21 (1.57) & 3.35 (1.84) & 3.34 (1.90) \\
        3 & 4 & 0 & 1.48 (0.73) & 2.38 (1.41) & 2.11 (1.46) & 3.26 (1.74) & 3.25 (1.80) \\ \hline
        4 & 1 & 1 & 2.42 (1.85) & 3.08 (2.22) & 2.49 (1.88) & 3.38 (2.18) & 3.32 (2.22) \\
        4 & 1 & 0 & 2.13 (1.74) & 3.17 (2.31) & 2.36 (1.82) & 3.15 (2.08) & 3.15 (2.14) \\
        4 & 2 & 1 & 2.08 (1.75) & 3.20 (2.33) & 2.54 (1.91) & 3.48 (2.22) & 3.41 (2.27) \\
        4 & 2 & 0 & 1.75 (1.49) & 2.81 (2.08) & 2.40 (1.82) & 3.34 (2.10) & 3.27 (2.15) \\
        4 & 3 & 1 & 2.00 (1.50) & 2.90 (2.12) & 2.60 (1.93) & 3.57 (2.25) & 3.53 (2.31) \\
        4 & 3 & 0 & 1.77 (1.20) & 3.03 (2.13) & 2.46 (1.82) & 3.48 (2.12) & 3.44 (2.17) \\ \bottomrule
    \end{tabularx}
\begin{tablenotes} \footnotesize
\item
\textit{Notes:}
Columns \textit{Team} and \textit{Opponent} indicate the respective seeding pots. Column \textit{Location} is a binary variable taking the value 1 if the team plays at home and 0 if it plays away.
\item
The tournament formats are detailed in Table~\ref{Table3}.
\item
The first value (second value in parenthesis) shows incentives for the first (second) prize based on Equation~\ref{eq:incentives}.
\end{tablenotes}
\end{threeparttable}
}
\end{table}

\end{appendices}

\end{document}